\documentclass[a4paper, article, english, oneside, unknownkeysallowed, 12pt]{memoir}
\usepackage{write-style}

\title{Inference on the TSLS Estimand \\ with Weak Instruments and \\ Treatment Effect Heterogeneity\thanks{Email: \href{mailto:arnsteinvestre@gmail.com}{arnsteinvestre@gmail.com}. I am grateful to Alex Torgovitsky, Magne Mogstad, Mikkel Plagborg-Møller, Adam McCloskey, Kirill Ponomarev, Christian Hansen and seminar participants at the Econometrics advising group at the University of Chicago for valuable feedback.}}

\author{
Arnstein Vestre \\[-4pt]
{\small University of Chicago} \\
}
\date{June, 2026
} %

\begin{document}

\maketitle

\begin{abstract}
\noindent
Traditional inference on the coefficient in an instrumental variables regression does not retain size when the instrument set is weak. With constant treatment effects or one instrument, the \cite{andersonEstimationParametersSingle1949} AR test, the \cite{kleibergenPivotalStatisticsTesting2002}--\cite{moreiraConditionalLikelihoodRatio2003} LM test, and the Moreira CLR test provide robust alternatives which retain validity.
Under treatment effect heterogeneity, no valid inference procedure exists in the overidentified setting. This paper develops the TSLS likelihood ratio (TLR) statistic,
for performing inference on the TSLS estimand.
When combined with a two-step procedure in the spirit of \cite{bergerValuesMaximizedConfidence1994}, it retains uniform validity across both the weak- and strong-instrument regimes.
The procedure retains power with small choices of first-step level, hence
the test can be constructed to
numerically coincide with the Wald test in the strong-instrument limit.
\end{abstract}

\clearpage
\section{Introduction}
\label{sec:introduction}

Consider the problem  of performing inference on a parameter of interest,
\begin{align}
    \mathcal{H}_0:\beta = \beta_0 \qquad\qquad \text{vs.} \qquad\qquad \mathcal{H}_1:\beta \neq \beta_0,
\end{align}
when this parameter is the coefficient in an overidentified instrumental variables regression with one endogenous regressor and a potentially weak set of instruments,  known as the two-stage least squares (TSLS) estimand, 
\begin{align}
    \quad  \beta \equiv \frac{\gamma'\Sigma_{ZZ}\delta}{\gamma'\Sigma_{ZZ}\gamma}.
    \label{eq:tsls-estimand}
\end{align}
The TSLS estimand is a weighted sum of the elements of the reduced-form coefficient vector, \(\delta\), weighted by the elements of the first-stage coefficient vector, \(\gamma\),
and the population instrument Gram matrix,  \(\Sigma_{ZZ}\).
It is a common parameter of interest in economics, political science, sociology, and other fields. Under certain assumptions, it obtains a causal interpretation as a positively weighted average of local average treatment effects, see e.g. \cite{angristTwoStageLeastSquares1995} and \cite{mogstadCausalInterpretationTwoStage2021}.

In the presence of weak instruments, the commonly applied Wald- or t-test is known to suffer from size distortion. 
This problem  is well-studied, and 
in the setting of a linear model with constant treatment effects (CTE),
the AR test (\citealp{andersonEstimationParametersSingle1949}), Kleibergen--Moreira Lagrange multiplier (LM) test (\citealp{kleibergenPivotalStatisticsTesting2002}, \citealp{moreiraConditionalLikelihoodRatio2003}) and conditional likelihood ratio (CLR) test (\citealp{moreiraConditionalLikelihoodRatio2003}) are known to deliver valid inference.\footnote{See  \cite{finlayImplementingWeakInstrumentRobust2009} and \cite{andrewsWeakInstrumentsInstrumental2019} for parsimonious expositions.}
In a model with
treatment effect heterogeneity, however, these are not valid tests for hypotheses on the TSLS estimand. 
As pointed out by \citet[p. 5]{yapInferenceManyWeak2025}, there exists no  test proven to be valid  (i) in the presence of weak instruments and (ii) treatment effect heterogeneity which (iii) targets a parameter interpretable as a weighted average of local average treatment effects under commonly maintained assumptions when (iv) considering limit experiments that  fix the number of instruments.

To understand why standard robust tests are not valid for hypotheses on the TSLS estimand under treatment effect heterogeneity, observe that
the AR test performs inference on the hypothesis \(\beta=\beta_0\) by instead performing inference on a set of linear restrictions,
\begin{align}
    \mathcal{H}_0: \delta-\beta_0\gamma= 0 \qquad \text{vs.} \qquad \mathcal{H}_1: \delta-\beta_0\gamma \neq 0.
\end{align}
where the number of restrictions is equal to the number of instruments.
The set of restrictions can be rewritten as a composite hypothesis on the following form,
\begin{align}
    \mathcal{H}_{0,(a)}: \beta=\beta_0 \qquad \text{and} \qquad \mathcal{H}_{0,(b)}:\delta-\beta\gamma= 0 \quad \text{for some } \ \beta\in\R.
\end{align}
As is shown, e.g., by \cite{moreiraConditionalLikelihoodRatio2003}, the AR test statistic can be written as the sum of the LM test statistic and the \cite{sarganEstimationEconomicRelationships1958}--\cite{hansenLargeSampleProperties1982} $J$ test statistic. Maintaining CTE, the $J$ test is a test of instrument exogeneity. Maintaining exogeneity, but allowing for treatment effect heterogeneity, the $J$ test is essentially a test of the CTE assumption.
 The CLR and LM tests are derived from the AR restrictions, testing  \(\beta=\beta_0\) while maintaining the assumption of CTE. As a consequence, they are
 invalid in the presence of treatment effect heterogeneity when considered as tests for the TSLS hypothesis.

This paper proposes a test that is valid with weak instruments and treatment effect heterogeneity.
The test is derived by first observing that a hypothesis about the TSLS estimand, \(\beta=\beta_0\), can be reformulated into a quadratic constraint by multiplying by the denominator on both sides and rearranging, 
\begin{align}
    \mathcal{H}_0:\gamma'\Sigma_{ZZ}(\delta -\beta_0\gamma) = 0
    \qquad \text{vs.} \qquad
    \mathcal{H}_1:\gamma'\Sigma_{ZZ}(\delta -\beta_0\gamma) \neq 0.
    \label{eq:quad-tsls-constraint}
\end{align}
While the restriction is similar in form to the AR restriction in that it does not involve division by a potentially small number,  it  is a single scalar restriction, and does not impose or test any ancillary assumptions, such as CTE.

The estimators of the reduced-form and first-stage coefficients are jointly asymptotically normal, with a variance-covariance matrix that is  consistently estimable.
This admits a likelihood ratio statistic, formulated as a quadratic optimization program with a quadratic constraint. We call this the TSLS likelihood ratio (TLR) statistic.
While the program does not have a closed-form solution, it is a special case of the generalized trust region subproblem, which has a known and unique solution, see e.g.
\cite{sternIndefiniteTrustRegion1995}.
   When instruments are strong, the TLR statistic is distributed asymptotically chi-squared with one degree of freedom. Combined with the fact that likelihood ratio tests are asymptotically equivalent to their associated Wald test under the null and against local alternatives (\citealp[p. 156]{ApproximationTheoremsMathematical1980}), this makes the implied test numerically indistinguishable from the Wald test in the strong-instrument regime.

When the instrument set is weak, the limiting distribution of the TLR statistic depends on two nuisance parameters.
The first nuisance parameter is a structural endogeneity parameter, capturing both selection on gains and levels. Similarly to the one-instrument case (see e.g. \citealp{vandesijpePowerConditionalLikelihood2023} and \citealp{leeWhatWhenYou2023}) it is consistently estimable under the null.
The second nuisance parameter captures the  overall identifying power of the instrument set. It is closely related to, but distinct from, the {concentration parameter} often used to gauge instrument strength in the literature. While it cannot be consistently estimated, it is associated with a sample statistic similar to, but distinct from, the first-stage F-statistic.
This allows for a two-step procedure  similar to the one discussed by \cite{bergerValuesMaximizedConfidence1994}, and suggested by \cite{staigerInstrumentalVariablesRegression1997}  for valid Wald inference in the linear constant effects model. Unlike their approach, we leverage more of the information contained in the sample to produce a test that is less conservative. In particular, we can set the first-step level so low as to numerically asymptote to the Wald test with strong instruments without substantial loss of power.

\paragraph*{Literature.}

The paper contributes to the literature on inference with weak instruments. Following the seminal contributions of \cite{nelsonDistributionInstrumentalVariables1990,nelsonFurtherResultsExact1990}, \cite{boundProblemsInstrumentalVariables1995} and \cite{staigerInstrumentalVariablesRegression1997}, a substantial number of papers have been devoted to understanding and resolving the finite-sample shortcomings of common procedures for inference and estimation with weak instruments. For a  comprehensive review, see \cite{andrewsWeakInstrumentsInstrumental2019}.
In the context of constant treatment effects,
\cite{andersonEstimationParametersSingle1949}
provided the first test with uniform validity, shown by \cite{moreiraTestsCorrectSize2009} 
to be
uniformly most powerful unbiased with one instrument.
Later, \cite{kleibergenPivotalStatisticsTesting2002}
 and \cite{moreiraConditionalLikelihoodRatio2003} developed the Kleibergen--Moreira LM and Moreira CLR tests, with better power properties.
\cite{andrewsOptimalTwoSidedInvariant2006} showed that the latter of these tests is uniformly most powerful unbiased under homoskedasticity.
Recently, progress has been made on deriving
more intuitive tests (\citealp{leeValidTRatioInference2022}) and  tests with more intentional direction of power (\citealp{leeWhatWhenYou2023}) in the one-instrument case, both  building on the
two-step correction procedure suggested by \cite{staigerInstrumentalVariablesRegression1997},
which uses the first-stage F-statistic and a Bonferroni correction to produce valid Wald inference.

The paper also contributes to the study  of instrumental variables in policy evaluation and models with heterogeneous treatment effects. For a review of this literature, see \cite{mogstadInstrumentalVariablesUnobserved2024}. While the TSLS estimand is one of many potential ways of aggregating reduced-form estimates when overidentified, following \cite{imbensIdentificationEstimationLocal1994b}, \cite{angristTwoStageLeastSquares1995} and \cite{angristInterpretationInstrumentalVariables2000}, it has been a common parameter of interest for empirical practitioners. Under certain assumptions, it obtains a weighted average of local average treatment effects. As is shown by \cite{kolesarEstimationInstrumentalVariables2013}, this is not always the case for estimands associated with other estimators in the literature, such as the LIML estimator introduced by \cite{andersonEstimationParametersSingle1949}, and the Fuller--$k$ estimator, introduced by \cite{fullerPropertiesModificationLimited1977}.
The methods considered in this paper are robust to violations of the so-called monotonicity (uniformity) assumption, discussed e.g. by \cite{heckmanUnderstandingInstrumentalVariables2006}, \cite{mogstadCausalInterpretationTwoStage2021} and \cite{sigstadMonotonicityJudgesEvidence2026}, as they do not rely on the causal interpretation of the estimand.
While the one-instrument setting is more widespread in the literature,
a common setting with more than one instrument is 
the examiner, judge or leniency design,
see \cite{chynExaminerJudgeDesigns2025} and \cite{goldsmith-pinkhamLeniencyDesignsOperators2025} for reviews. In terms of valid inference in the presence of heterogeneous treatment effects, \cite{evdokimovInferenceInstrumentalVariables2018} study the problem in a setting with many instruments, and \cite{yapInferenceManyWeak2025} with many weak instruments. To the author's knowledge, this is the first paper to consider the problem of valid inference on the TSLS estimand with a fixed number of weak instruments and potentially heterogeneous treatment effects.

The paper is also related to the literature using two-step procedures to obtain valid inference in the presence of nuisance parameters, see e.g. \cite{lohNewMethodTesting1985},
\cite{bergerValuesMaximizedConfidence1994}, \cite{silvapulleTestPresenceNuisance1996}, \cite{hansenTestSuperiorPredictive2005}, \cite{chernozhukovIntersectionBoundsEstimation2013}, \cite{romanoPracticalTwoStepMethod2014} and \cite{mccloskeyBonferronibasedSizecorrectionNonstandard2017}. The two-step test proposed by \cite{staigerInstrumentalVariablesRegression1997} is an example of such a testing procedure in the weak-instrument setting.

\paragraph*{Roadmap.}
The paper proceeds as follows.
In \Cref{sec:set-up-not} we set up the linear instrumental variables model with heterogeneous treatment effects and derive a  statistic that captures the total identifying power of the instrument set. In \Cref{sec:ar-clr} we show that existing weak-instrument robust tests fail as tests for the TSLS hypothesis with  treatment effect heterogeneity. In
\Cref{sec:results} we introduce the TLR statistic, and derive its limiting distribution. We show that it retains validity when combined with a two-step pretest and study its power properties in the weak- and strong-instrument regime. \Cref{sec:conclusion} concludes.

\section{Set-up and Notation}

\label{sec:set-up-not}

\paragraph*{Fundamentals.}
Let \(\{(Y_i,D_i,Z_i)\}_{i=1}^n\) be an i.i.d. sample, where \(Y_i\) is an outcome, \(D_i\) a treatment, both scalar, and \(Z_i\) a \(d_Z\)-dimensional vector of instruments, with \(d_Z\geq 2\).
Assume  all variables are demeaned and have finite fourth moments.
In the presence of covariates, let \(Y_i\) and \(D_i\) denote variables where their impact is appropriately projected off.
We are interested in the linear  heterogeneous treatment effects model,
\begin{align}
       Y_i &= \beta_i D_i + U_i\label{eq:model-equation}
\end{align}
where \(\beta_i\) is the causal effect of \(D_i\) on \(Y_i\), and \(U_i\) is the model residual.
Denote the linear regressions of  \(Y_i\) on \(Z_i\)  (reduced form) and \(D_i\) on \(Z_i\)  (first stage) as,
\begin{align}
Y_i &=  \delta_n' Z_i + W_i \label{eq:reduced-form-equation}\\
D_i &= \gamma_n' Z_i + V_i \label{eq:first-stage-equation}
\end{align}
where \(\delta_n\) is the vector of reduced-form coefficients, \(\gamma_n\) the vector of first-stage coefficients, and \(W_i\) and \(V_i\)
are regression residuals.
Subscripts on \(\delta_n\) and \(\gamma_n\) indicate that we do not restrict
the data generating process to be invariant with respect to the sample size \(n\).
To analyze weak-instrument properties, we follow \cite{staigerInstrumentalVariablesRegression1997} in letting \((\delta_n, \gamma_n)\) drift towards zero at rate \(1/\sqrt{n}\), such that
\((\delta_n,\gamma_n) \equiv  { (\delta,\gamma)}/{\sqrt{n}}\) for some vectors  \((\delta, \gamma)\).
To study strong-instruments properties, we hold \((\delta_n, \gamma_n)\) fixed as \(n\to\infty\). Let the TSLS estimand, \(\beta\), be defined as in \cref{eq:tsls-estimand}. It is uniquely pinned down by  \((\delta, \gamma)\), hence the parameter of interest does not vary with the sample size. 
Let \(D_i(z)\) denote the potential treatment of an individual with instrument realization \(Z_i=z\).
We maintain the following assumptions.
\begin{assumptionp}{IVR}[Relevance]\label{ass:ivr}
\(\E[Z_iD_i] \neq \mathbf{0}_{d_Z}\)
\end{assumptionp}

\begin{assumptionp}{IVX}[Strong Exogeneity]\label{ass:ivx}
\((\beta_i,U_i,\{D_i(z)\}_{z\in\R^{d_Z}} ) \indep Z_i\)
\end{assumptionp}

\noindent
Assumptions \ref{ass:ivr} and \ref{ass:ivx}
are a subset of the assumptions usually maintained
in models with heterogeneous treatment effects, see e.g. \cite{angristTwoStageLeastSquares1995} and \cite{angristInterpretationInstrumentalVariables2000}.
No assumption is imposed on  the monotonicity (uniformity) of \(D_i(z)\) in \(z\). However, treatment effect heterogeneity requires the following refinement to the \cite{staigerInstrumentalVariablesRegression1997} weak-instrument limiting sequence.
\begin{assumptionp}{IVW}[Uniform Weakness]\label{ass:ivw}
For \(\sqrt{n}(\delta_n,\gamma_n)=(\delta,\gamma)\), as  \(n\to\infty\) and
\begin{enumerate}[label={\emph{(\alph*)}}, ref={(\alph*)}]
    \item  \label{ass:ivw-a}
    \(D_i\) binary, \(\lim_{n\to\infty}\sup_{z,z'}\P\bigl[D_i(z)\neq D_i(z')\bigr]=0\)
    \item  \label{ass:ivw-b}
    \(D_i\) continuous, \(\lim_{n\to\infty}\sup_{z,z'}\E\bigl[(D_i(z)- D_i(z'))^2\bigr]=0\)
\end{enumerate}
\end{assumptionp}

\noindent
Assumption \ref{ass:ivw} rules out the case in which an instrument is weak because the share of individuals who are  instrumented into more treatment (compliers) and less treatment (defiers) cancel. It is implied by the assumption of monotonicity (uniformity) of \(D_i(z)\) in \(z\), or equivalently by the \cite{vytlacilIndependenceMonotonicityLatent2002a} single-index model.

\paragraph*{Observed moments.}
Let \(\mathfrak{N}(\,\cdot\,,\,\cdot\,)\) denote the Normal distribution, \(\tau_n \equiv \left[\begin{smallmatrix}
    \delta_n \\
    \gamma_n
\end{smallmatrix}\right]\)
  the stacked vector of first-stage and reduced-form coefficients and \(\hat \tau\equiv \left[\begin{smallmatrix}
    \hat \delta \\
    \hat \gamma
\end{smallmatrix}\right]\)
 its OLS estimator. While the estimator is also a function of the sample size, for brevity of notation we let this be implicit in the following. 
 The estimator has the following limiting distribution.
\begin{align}
    \sqrt{n}(\hat\tau-\tau_n) \convd \mathfrak{N}(\mathbf{0}_{2d_Z},\Sigma),
    \qquad\quad
    \Sigma   \equiv
   \begin{bmatrix}
       \Sigma_{\hat \delta} & \Sigma_{\hat \delta,\hat \gamma} \\
       \Sigma_{\hat \delta,\hat \gamma}' & \Sigma_{\hat \gamma}
   \end{bmatrix}
\end{align}
Let \(\Sigma_{ZZ}\equiv \E[Z_iZ_i']\) denote the population Gram matrix of \(Z_i\).
We can rewrite
the constraint defining the TSLS hypothesis from \cref{eq:quad-tsls-constraint} in terms of \(\tau_n\) as follows.
\begin{align}
    \tau_n'\,\Gamma(\beta_0) \,\tau_n = 0 \qquad \text{where} \qquad
    \Gamma(\beta_0) \equiv
    \begin{bmatrix}
        0 & \Sigma_{ZZ} \\
        \Sigma_{ZZ} & -2\beta_0\Sigma_{ZZ}
    \end{bmatrix}
\end{align}
\noindent 
With constant treatment effects, the first-stage F-statistic summarizes the strength of the instrument set. With treatment effect heterogeneity, we need to consider a statistic that also captures other facets of the identifying power of the instruments. We record this statistic and its limiting distribution as a lemma.

\begin{lemmap}{ST}[Summary Statistic]\label{lem:st}
Let \(\hat \Sigma\) denote a consistent estimator for \(\Sigma\), and let \(\hat S \equiv \lVert (\hat\Sigma/n)^{-\frac{1}{2}}\hat \tau\rVert^2/d_Z\) denote the Wald test statistic for the  hypothesis \(\tau_n=0\). Let \(\mathfrak{C}^2_{2d_Z}(\xi)\) denote the non-central chi-squared distribution with \(2d_Z\) degrees of freedom and non-centrality parameter \(\xi\). Then, {for \(\sqrt{n}\,\tau_n = \tau\), as \(n\to\infty\),}
    \[d_Z\hat S  \convd \mathfrak{C}^2_{2d_Z}\left(\xi\right),\]
where \(\xi \equiv \lVert (\Sigma/n)^{-\frac{1}{2}}\tau_n\rVert^2\).
\end{lemmap}

\begin{proof}
    See Appendix~\ref{sec:proof-st}.
\end{proof}
\noindent
The parameter \(\xi\) is related to the concentration parameter often used to enumerate instrument strength in the literature. We record the relation as a lemma.
\begin{lemmap}{PD}[Parameter Decomposition]\label{lem:pd}
Let \(\mu^2\equiv \lVert  (\Sigma_{\hat \gamma}/n)^{-\frac{1}{2}}\gamma_n\rVert^2\) denote the concentration parameter, 
\(\omega\equiv \beta_{\mathrm{ols}}-\beta\), \(\beta_{\mathrm{ols}}\equiv \cov[Y_i,D_i]/\var[D_i]\), the difference between the OLS and TSLS estimands, and 
\(\nu^2 \equiv \lVert \Sigma_{ZZ}^{{1}/{2}}(\delta_n - \beta \gamma_n)\rVert^2 / \lVert \Sigma_{ZZ}^{{1}/{2}} \gamma_n\rVert^2\)
the TSLS-weighted dispersion of
local average treatment effects.
Under Assumptions \ref{ass:ivx} and \ref{ass:ivw}, as \({\sqrt{n}\,\tau_n=\tau}\) and \(n\to\infty\), we have,
\begin{align*}
    \xi = \mu^2\left(1 + h^{-2}(\nu^2 + \omega^2)\right), 
\end{align*}
where 
\(h^2\equiv{(\var[W_i]-\beta_{\mathrm{ols}}^{2}\var[V_i])/\var[V_i]}\) is a  ratio of residual variances.
All parameters are constant along the weak-instrument asymptotic sequence.
\end{lemmap}

\begin{proof}
    See Appendix~\ref{sec:proof-pd}.
\end{proof}
\noindent 
Observe that with one instrument or constant treatment effects, we have \(\nu^2=0\).

\paragraph*{Eigenvalue Representation.}

It will be useful to express the data in terms of
a particular eigendecomposition. 
Let \(\Lambda(\beta_0) \equiv
\Sigma^{\frac{1}{2}}
\Gamma(\beta_0)
\Sigma^{\frac{1}{2}}\) denote
 the rotation of  \(\Gamma(\beta_0)\) in terms of \(\Sigma\).
Let \(X\) denote the matrix of eigenvectors of \(\Lambda(\beta_0)\), and \(\hat X\) its analog estimator.
 Define the rotated first-stage and reduced-form estimators and estimands as
\begin{align}
    \hat Q \equiv \hat X'\hat\Sigma^{-1/2}\sqrt{n}\,\hat\tau
    \qquad\text{and}\qquad
    q \equiv X'\Sigma^{-1/2}\sqrt{n}\,\tau_n.
\end{align}
While
\(\Lambda(\beta_0)\) is indefinite, the block-structure of \(\Gamma(\beta_0)\) and \Cref{ass:ivx} restrict the sign of, and variation in, its eigenvalues.
\begin{lemmap}{EH}[Eigenvalue Homogeneity]\label{lem:eh}
\(\Lambda(\beta_0)\) has \(d_Z\) positive and \(d_Z\) negative eigenvalues.
Let \(\{\kappa^{(k)}_-,\kappa^{(k)}_+\}_{k=1}^{d_Z}\) denote these eigenvalues, and let \((q_+,q_-)\) denote the subvectors of \(q\) associated with the positive and negative eigenspace. 
Under Assumptions \ref{ass:ivx} and \ref{ass:ivw}, along any sequence with \(\sqrt{n}\,\tau_n=\tau\), as \(n\to\infty\), for some common \((\kappa_-,\kappa_+)\), we have
    \begin{enumerate}
    \item 
        \(\kappa^{(k)}_-\to\kappa_-\) and  \(\kappa^{(k)}_+\to\kappa_+\)  for all \(k=1,\ldots,d_Z\).
    \item 
    \(\tau_n'\,\Gamma(\beta_0) \,\tau_n = 0 \iff (1+\varrho)\lVert q_+ \rVert^2 = (1-\varrho)\lVert q_- \rVert^2\)
    \end{enumerate}
    where \(\varrho \equiv  (\kappa_+-\lvert \kappa_-\rvert) / (\kappa_++\lvert\kappa_-\rvert)\) denotes the signed eigenvalue share. If we let \(\omega_0 \equiv \beta_{\mathrm{ols}} - \beta_0 \) denote the value \(\omega\) takes under the null, we have \(\varrho = \omega_0 /\sqrt{h^2+\omega_0^2}\).
\end{lemmap}

\begin{proof}
    See Appendix~\ref{sec:proof-eh}.
\end{proof}

\noindent
Note that \((q, X, \kappa_\pm, \varrho)\) and their estimators depend on the hypothesized \(\beta_0\). For brevity of notation, we suppress this in the following.

\section{On the Validity of Weak-Instrument Robust Tests}

\label{sec:ar-clr}

The AR, LM and CLR tests are commonly understood to provide inference that is robust to weak instruments. In the following, we show that this  is not the case if the tests are interpreted as tests of the TSLS estimand under treatment effect heterogeneity.
We start by recording the constant treatment effect limiting distribution of the three statistics, as derived by \cite{andersonEstimationParametersSingle1949}, \cite{kleibergenPivotalStatisticsTesting2002} and \cite{moreiraConditionalLikelihoodRatio2003}.

\begin{lemmap}{ROB-CTE}[Robustness, Constant Treatment Effects]\label{lem:rob-cte}
Assume there exists a scalar \(\beta\) such that \(\delta_n=\beta\gamma_n\).
Let \(\hat\Omega(\beta_0)\) denote a consistent estimator of the variance of the restriction \(\hat\delta - \beta_0\hat\gamma\), let \(\tilde\gamma(\beta_0)\)
denote the first-stage estimator orthogonalized relative to \(\hat\delta - \beta_0\hat\gamma\) and let \(\hat\Psi(\beta_0)\) denote an estimator of its variance.
Define
\begin{align*}
\mathrm{AR}(\beta_0)  &\equiv d_Z^{-1}\lVert (\hat\Omega(\beta_0)/n)^{-1/2}(\hat\delta - \beta_0\hat\gamma)\rVert^2,   \\[2pt]
\mathrm{LM}(\beta_0)  &\equiv d_Z^{-1}\bigl[\tilde\gamma(\beta_0)'(\hat\Omega(\beta_0)/n)^{-1}(\hat\delta - \beta_0\hat\gamma)\bigr]^2\,\big/\,\lVert(\hat\Omega(\beta_0)/n)^{-1/2}\tilde\gamma(\beta_0)\rVert^2,\\[2pt]
\mathrm{CLR}(\beta_0) &\equiv \tfrac{1}{2}\!\left[d_Z\mathrm{AR}(\beta_0)- \lVert\hat r(\beta_0)\rVert^2 + \sqrt{(d_Z\mathrm{AR}(\beta_0)-\lVert\hat r(\beta_0)\rVert^2)^2 + 4 \,d_Z\mathrm{LM}(\beta_0)\, \lVert\hat r(\beta_0)\rVert^2} \right]
\end{align*}
where \(\hat r(\beta_0) \equiv (\hat\Psi(\beta_0)/n)^{-1/2}\,\tilde\gamma(\beta_0)\).
Let \(\mathfrak{C}^2_{d_Z}\) denote the chi-squared distribution with \(d_Z\) degrees of freedom. We then have, \(d_Z\mathrm{AR}(\beta_0)\convd \mathfrak{C}^2_{d_Z}\), \(d_Z\mathrm{LM}(\beta_0)\convd \mathfrak{C}^2_1\) and \(\mathrm{CLR}(\beta_0)\convd m(\mathfrak{C}^2_{1},\mathfrak{C}^2_{d_Z-1},\lVert \hat r(\beta_0)\rVert^2)\), where
\begin{align*}
m(x_1,x_{d_Z-1},r) \;\equiv\;
\tfrac{1}{2}\!\left[x_{d_Z-1}+x_1- r + \sqrt{(x_{d_Z-1}+x_1-r)^2 + 4\, x_1\, r} \right].
\end{align*}
\end{lemmap}

\begin{proof}
    See Appendix~\ref{sec:proof-rob-cte}.
\end{proof}
\noindent
With treatment effect heterogeneity, these limits no longer obtain. We record the result as a lemma.
\begin{lemmap}{ROB-HTE}[Robustness, Heterogeneous Treatment Effects]\label{lem:rob-hte}
Let \(\beta=\beta_0\) denote the TSLS hypothesis. 
Let \(\theta_{\mathrm{AR}} \equiv \mu^2\nu^2/(h^2+\omega^2)\), and let
 \(\zeta\sim \mathfrak{N}(0,1)\),
\(\zeta^2_{1} \sim \mathfrak{C}^2_{1}(\omega^2\theta_{\mathrm{AR}}/h^2)\), and \(\zeta^2_{d_Z-1} \sim \mathfrak{C}^2_{d_Z-1}(\mu^2(h^2+\omega^2)/h^2)\) denote three random variables, where \(\zeta \indep \zeta^2_1 \indep \zeta^2_{d_Z-1}\).
Then, as \(\sqrt{n}\tau_n=\tau\), \(n\to\infty\),
\begin{enumerate}
    \item
    \(d_Z\mathrm{AR}(\beta_0)\convd \mathfrak{C}^2_{d_Z}(\theta_{\mathrm{AR}})\)
    \item
    \(d_Z\mathrm{LM}(\beta_0)\convd (\Upsilon+\zeta)^2\), with
    \( \Upsilon^2 \;\sim\; \theta_{\mathrm{AR}}\cdot{\zeta^2_1}/({\zeta^2_1 + \zeta^2_{d_Z-1}})\)

    \item
    \(\mathrm{CLR}(\beta_0) \convd m\bigl((\Upsilon+\zeta)^2,\;\mathfrak{C}^2_{d_Z}(\theta_{\mathrm{AR}}) - (\Upsilon+\zeta)^2,\;\lVert \hat r(\beta_0)\rVert^2 \bigr)\)
\end{enumerate}
Let \(c^T_{1-\alpha}\) denote the \((1-\alpha)^{th}\) quantile of the constant-treatment-effects limiting distribution of test statistic \(T\) and assume \(\omega^2 >0\). Then, for \(\nu^2>0\),
\[
\limsup_{n\to\infty}\,\P\left[\mathrm{T}(\beta_0)> c^{\mathrm{T}}_{1-\alpha}\right] > \alpha
\]
where \(T\in\{\mathrm{AR},\mathrm{LM},\mathrm{CLR}\}\). Further, \(\lim_{\mu^2\to\infty}\,\limsup_{n\to\infty}\,\P[\mathrm{T}(\beta_0)> c^{\mathrm{T}}_{1-\alpha}] =1\).
\end{lemmap}

\begin{figure}[!htb]
    \centering
    \subbottom[\(\nu^2=0\), High endogeneity]{\includegraphics[width=0.49\textwidth]{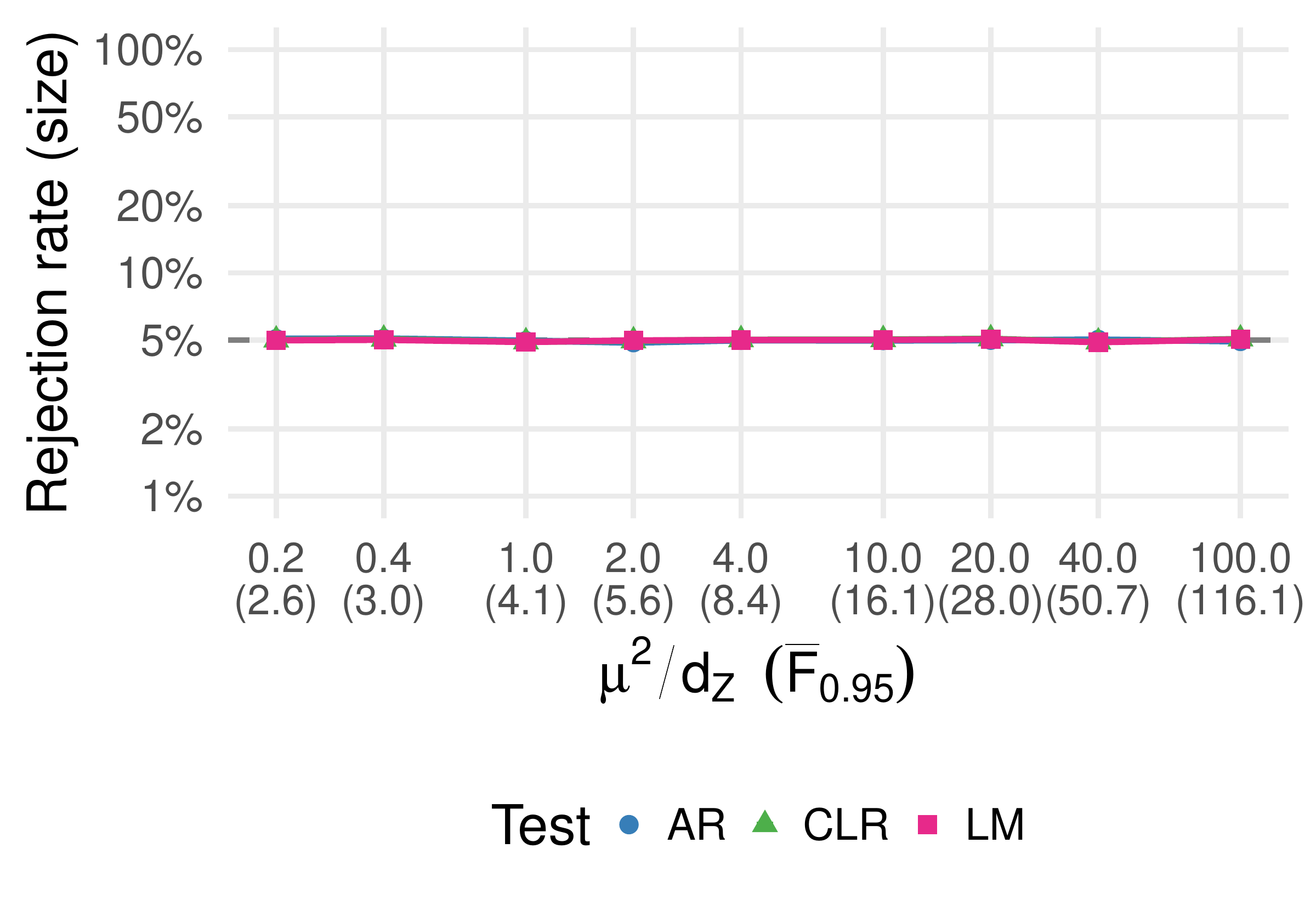}}
    \subbottom[\(\nu^2>0\), High endogeneity]{\includegraphics[width=0.49\textwidth]{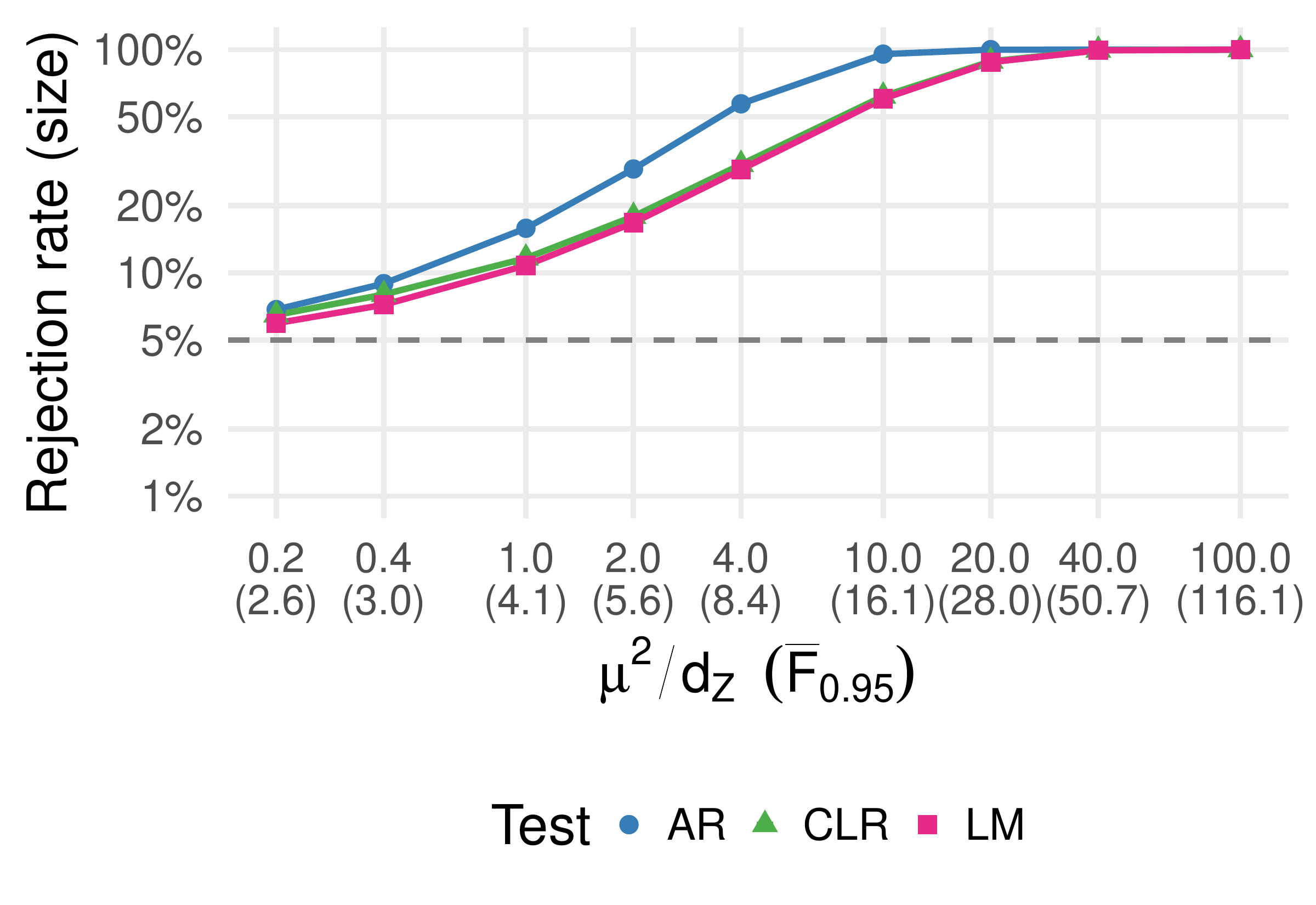}}
    \subbottom[\(\nu^2=0\), Intermediate endogeneity]{\includegraphics[width=0.49\textwidth]{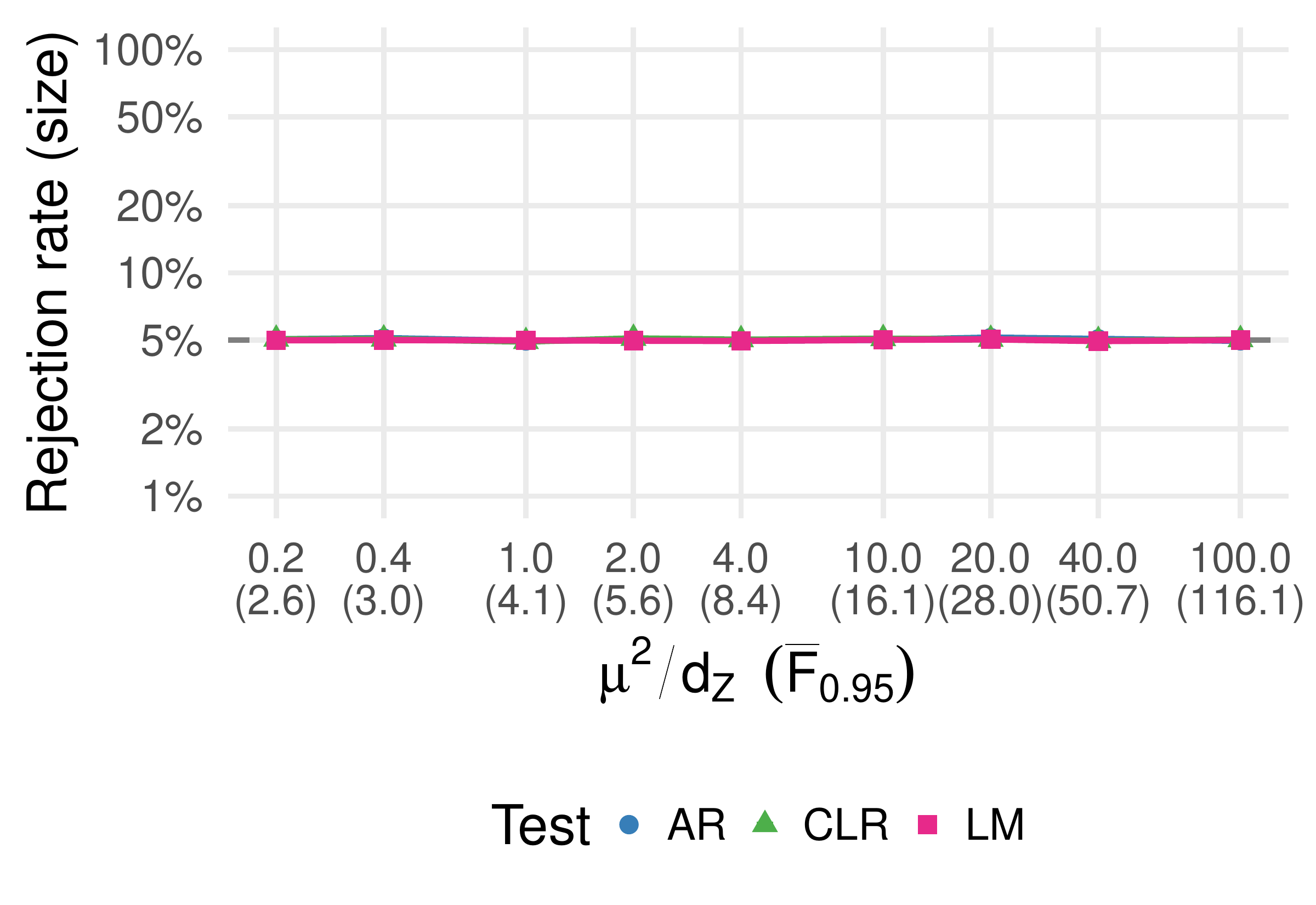}}
    \subbottom[\(\nu^2>0\), Intermediate endogeneity]{\includegraphics[width=0.49\textwidth]{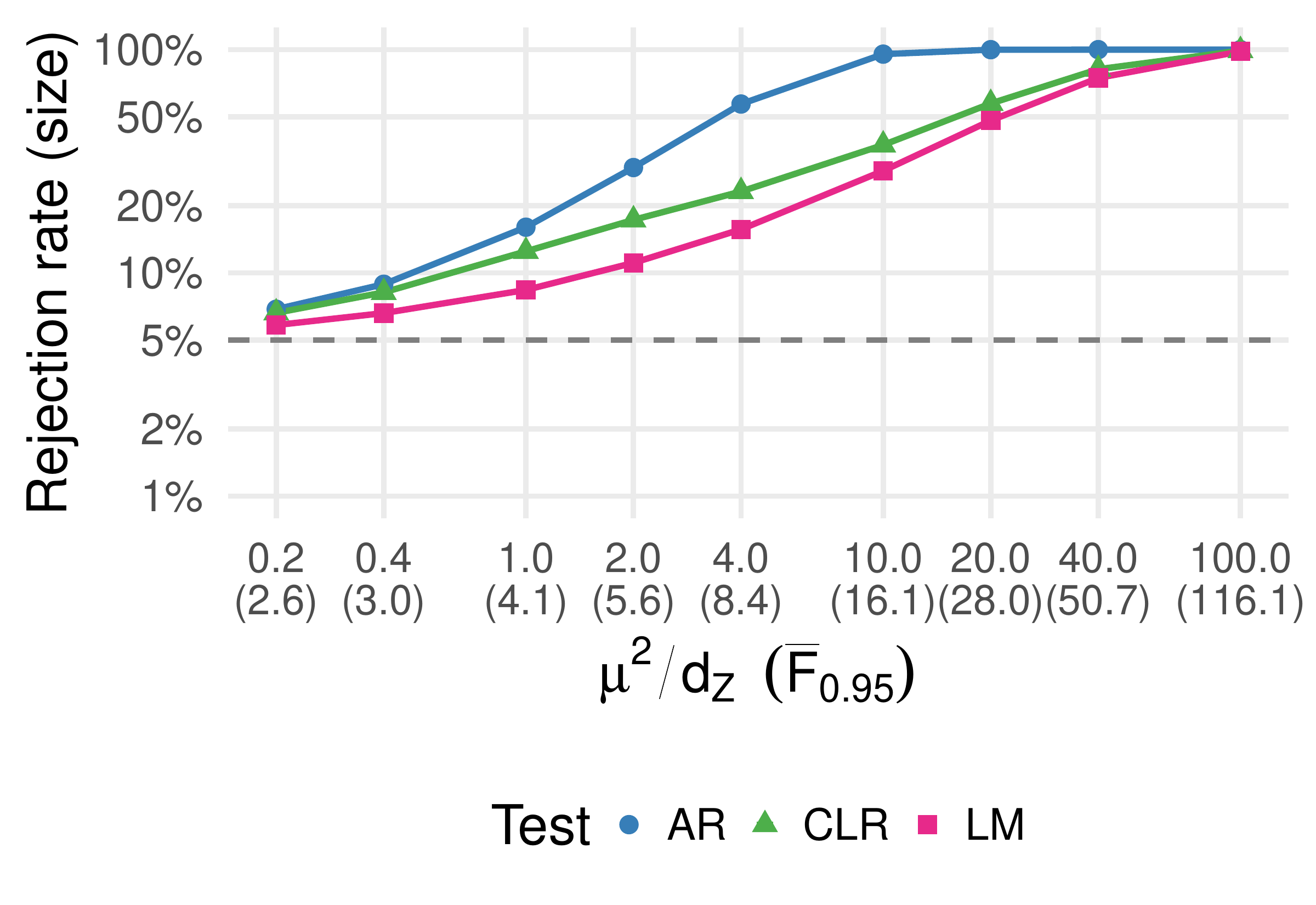}}
    \subbottom[\(\nu^2=0\), No endogeneity]{\includegraphics[width=0.49\textwidth]{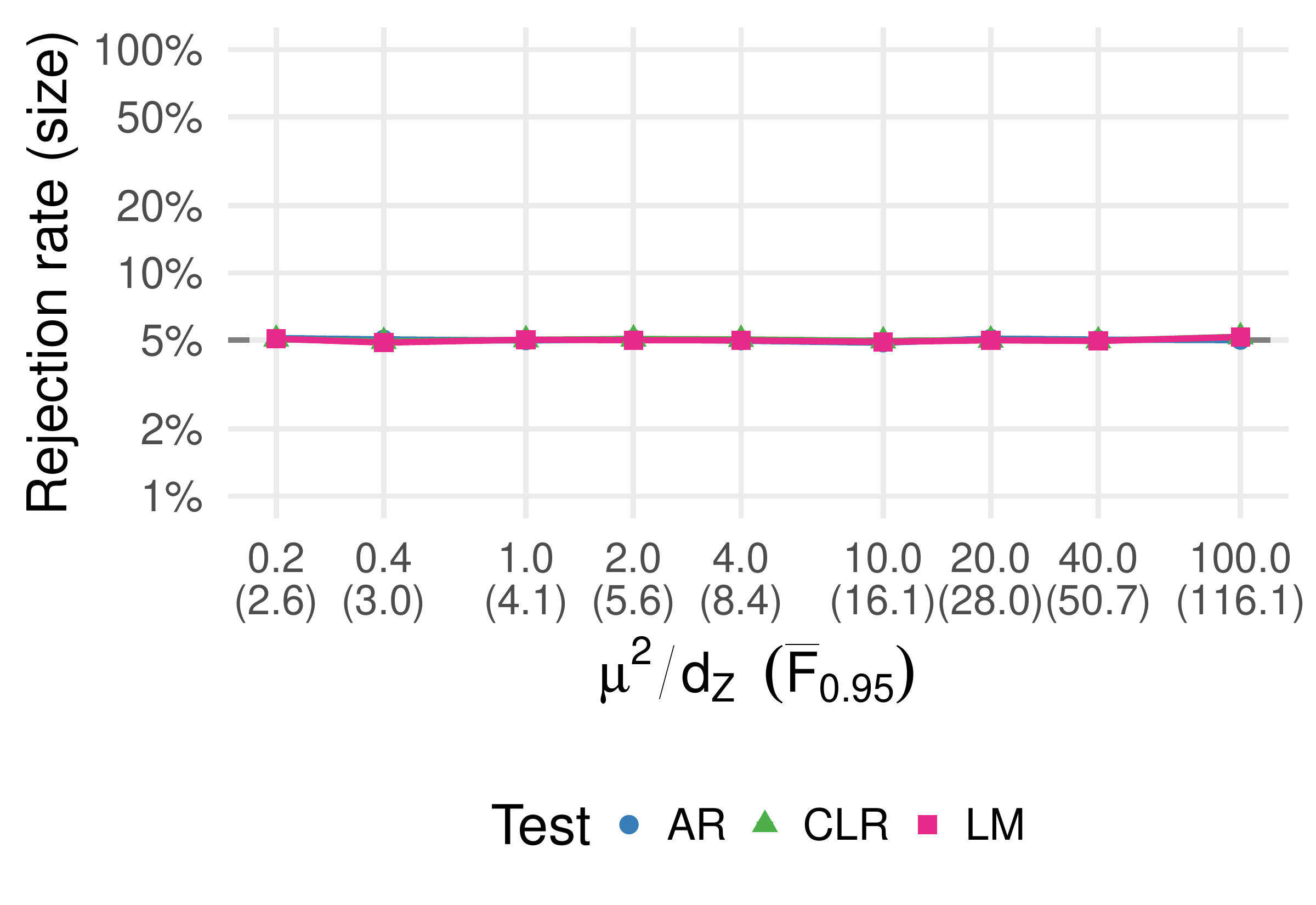}}
    \subbottom[\(\nu^2>0\), No endogeneity]{\includegraphics[width=0.49\textwidth]{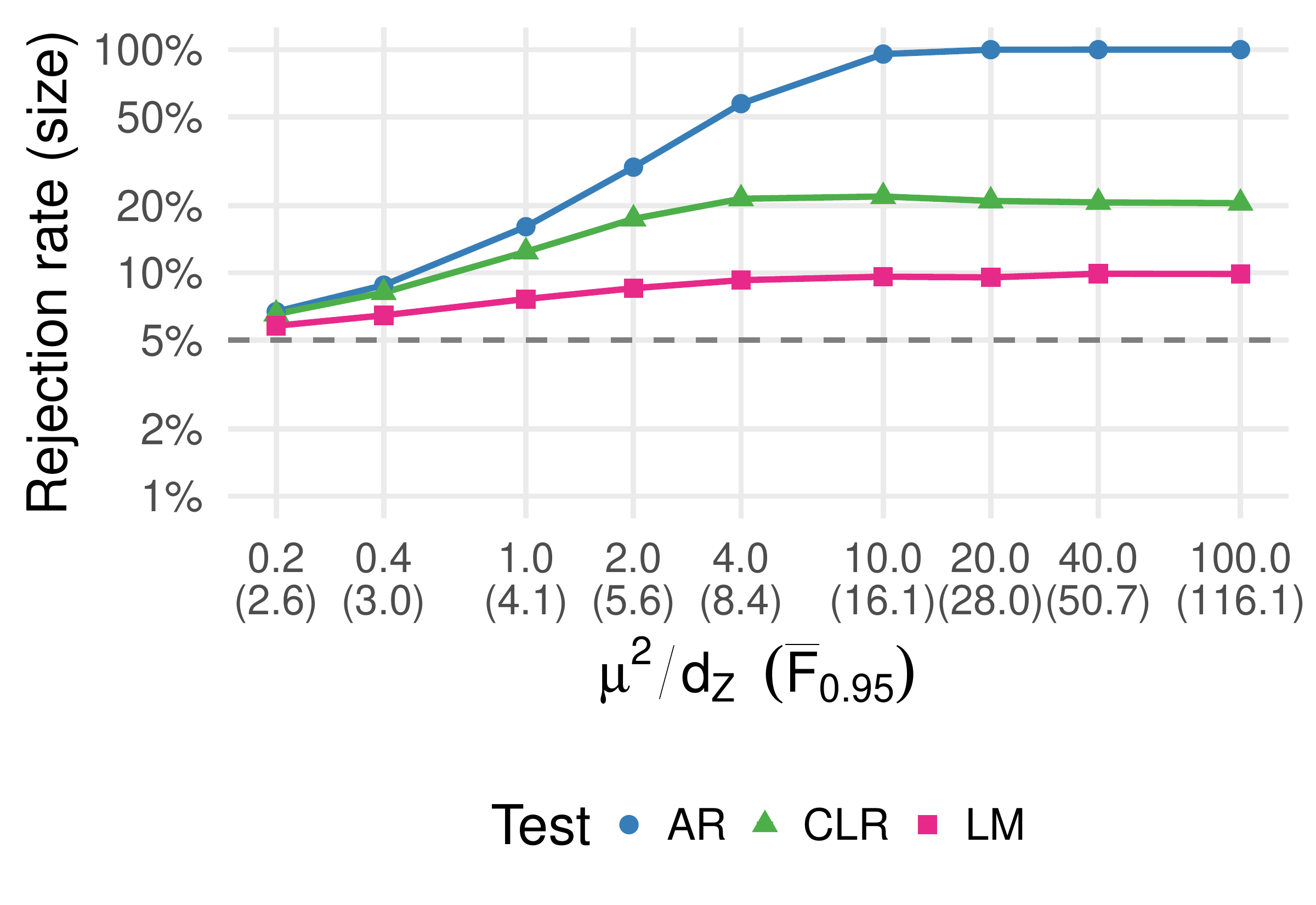}}

    \caption{\textbf{Size distortion of the AR, LM and CLR Tests Under Heterogeneous Treatment Effects.} This figure plots the size of the AR, LM and CLR tests if taken as tests of the hypothesis \(\beta=\beta_0\), under constant (\(\nu^2=0\)) and heterogeneous (\(\nu^2>0\)) treatment effects, with \(d_Z=5\) instruments.
    The blue line shows the AR test, the red LM, and the green the CLR test.
    Size is computed by simulating asymptotic realizations of \(\hat \tau\) under high, intermediate and no endogeneity, corresponding to \(\omega/\sqrt{h^2+\omega^2} \in \{0.9, 0.5, 0\}\).
    The y-axis shows rejection rate under the null (size), and the x-axis enumerates instrument strength in terms of the concentration parameter and the \(95^{\mathrm{th}}\) quantile of the associated non-central chi-squared distribution.}
    \label{fig:other-tests}
\end{figure}

\begin{proof}
    See Appendix~\ref{sec:proof-rob-hte}.
\end{proof}
\noindent
In \Cref{fig:other-tests}, we plot the size of the AR, LM and CLR tests, 
if used as tests for the TSLS hypothesis, \(\beta=\beta_0\), in a Monte Carlo simulation, with and without treatment effect heterogeneity. We simulate by drawing realizations of \(\hat \tau\) from its limiting distribution, with known variance,  consistent with constant treatment effects (panels (a), (c) and (e)), and heterogeneous treatment effects (panels (b), (d) and (f)), and compute the rejection rate  under the null. The x-axis is enumerated in terms of the \(d_Z\)-scaled concentration parameter, \(\mu^2/d_Z\). The associated first-stage F-statistic cutoffs that reject this concentration parameter at level \(\alpha=0.05\) are indicated in parentheses. 

In panels (a), (c) and (e) of \Cref{fig:other-tests}, we see that all three tests retain size under constant treatment effects, irrespective of the level of endogeneity. In panels (b), (d) and (f), we see that under heterogeneous treatment effects, the tests reject more often than their nominal level indicates, and are thus invalid as tests for the TSLS hypothesis, with size distortion increasing in instrument strength. With sufficiently strong instruments and non-zero endogeneity, the tests always reject.

\section{Valid Inference on the TSLS Estimand with Weak Instruments}
\label{sec:results}

We now turn to deriving a test for the TSLS hypothesis that is valid with weak instruments and treatment effect heterogeneity. We start by deriving the likelihood ratio statistic for the TSLS hypothesis. We then proceed to derive its strong- and weak-instrument limiting distributions. We conclude by showing  that we can perform uniformly valid inference by combining the test with a pretest for the nuisance parameter in a two-step procedure without meaningful loss of power.

\paragraph*{The TLR Statistic.}

Consider the asymptotic log-likelihood of \(\tau_n\) given \(\hat \tau\), given as,
\begin{align}
    \ell(\tau_n\mid \hat \tau) 
    = 
    L - \tfrac{1}{2}\,n(\hat \tau -\tau_n)'\hat\Sigma^{-1}(\hat \tau - \tau_n)
\end{align}
where \(L\) is a constant. 
Unconstrained, the log-likelihood is
maximized by \(\hat \tau\). Under the TSLS constraint, the maximized log-likelihood is a quadratic program with a quadratic constraint. Considering their difference, we get the likelihood ratio statistic for the TSLS hypothesis.
The implied quadratic program has a known solution,  due to \cite{sternIndefiniteTrustRegion1995}.
We record the result as a proposition.

\begin{propositionp}{TLR}[TSLS Likelihood Ratio Statistic]\label{prop:tlr}
The asymptotic likelihood ratio statistic for the hypothesis \(\beta=\beta_0\), when scaled by \(d^{-1}_Z\), is given as
\begin{align*}
\mathrm{TLR}(\beta_0) \equiv \min_{\tau_n \in \R^{2d_Z}}\,d_Z^{-1}n(\hat \tau -\tau_n)'\hat\Sigma^{-1}(\hat \tau - \tau_n) \qquad \text{s.t.} \qquad \tau_n'\,\hat\Gamma(\beta_0) \,\tau_n = 0
\end{align*}
where \(\hat\Gamma(\beta_0)\) is a consistent estimator for \(\Gamma(\beta_0)\).
Let \(\hat \Lambda(\beta_0)\) denote a consistent estimator for \(\Lambda(\beta_0)\).
The program has a unique solution,
    \begin{align*}
    \mathrm{TLR}(\beta_0) = d_Z^{-1}\sum_{j=1}^{2d_Z} \hat Q_j^2\!\left(\frac{\lambda^*\hat\kappa_j}{1+\lambda^*\hat\kappa_j}\right)^{\!2}
    \quad\text{s.t.}\quad
    \sum_{j=1}^{2d_Z}\frac{\hat\kappa_j \hat Q_j^2}{(1+\lambda^*\hat\kappa_j)^2} = 0
\end{align*}
where \(\{\hat \kappa_j\}_{j=1}^{2d_Z}\) denote the eigenvalues of \(\hat \Lambda(\beta_0)\), \(\hat Q_j\) the \(j^{\mathrm{th}}\) element of \(\hat Q\) and \(\lambda^*\in(-(\max_{j:\hat\kappa_j>0}\hat\kappa_j)^{-1},-(\min_{j:\hat\kappa_j<0}\hat\kappa_j)^{-1})\) is implicitly defined and unique on this interval.
\end{propositionp}

\begin{proof}
    See Appendix~\ref{sec:proof-tlr}.
\end{proof}

\noindent
We scale by \(d_Z^{-1}\) to simplify comparison between settings with different numbers of instruments.
Observe that the above result only relies on the joint asymptotic normality of the reduced form and first stage, and does not require any further assumptions.

Under maintained assumptions,
the TLR statistic has a limiting distribution that is well-behaved with strong instruments and depends  on two nuisance parameters when the instruments are weak. We record the result.

\begin{propositionp}{LD}[TLR Limiting Distribution]\label{prop:ld}
Let \(S\) be defined such that \(d_Z\hat S \convd S\).
    Under Assumptions \ref{ass:ivr}, \ref{ass:ivx} and \ref{ass:ivw}, the TLR statistic obtains the following limiting distribution.
    \begin{enumerate}
        \item
        When \(\tau_n\) is fixed and \(n\to\infty\),
        \(d_Z\,\mathrm{TLR}(\beta_0)\convd\mathfrak{C}^2_1\).
        \item
        When \(\sqrt n\,\tau_n=\tau\) is fixed and \(n\to\infty\),
        \[
            d_Z\,\mathrm{TLR}(\beta_0)\;\convd\;\tfrac{1}{2}\bigl(\sqrt{(1+\varrho)\,S_+}-\sqrt{(1-\varrho)\,S_-}\bigr)^{\!2},
        \]
        where \(S_+\sim\mathfrak{C}^2_{d_Z}((1-\varrho)\,\xi/2)\), \(S_-\sim\mathfrak{C}^2_{d_Z}((1+\varrho)\,\xi/2)\), \(S_+ \indep S_-\), and \(S_+ + S_-=S\).
    \end{enumerate}
\end{propositionp}

\begin{proof}
    See Appendix~\ref{sec:proof-ld}.
\end{proof}
\noindent
The weak-instrument limiting distribution  obtains the strong-instrument limit as \(\xi\to \infty\). We record this, and two other useful boundary results, as a proposition.
\begin{propositionp}{BD}[Boundary Case Distributions]\label{prop:bd}
Consider the weak-instrument regime, such that \(\tau_n\sqrt{n}=\tau\) constant as \(n\to\infty\). Then,
\begin{enumerate}
    \item
    As \(\xi\to \infty\), we have
    \(d_Z\mathrm{TLR}(\beta_0) \convd
    \mathfrak{C}^2_1\).
    \item
    As \(\varrho\to \pm 1\), we have
    \(d_Z\mathrm{TLR}(\beta_0) \convd
     \mathfrak{C}^2_{d_Z}\).
    \item
    At  \(\xi= 0\), \(\varrho=0\), we have
    \(d_Z\mathrm{TLR}(\beta_0) \convd
    \left(\sqrt{\mathfrak{C}^2_{d_Z}} -\sqrt{\mathfrak{C}^2_{d_Z}}\right)^2 \,\big/\,2\).
\end{enumerate}
\end{propositionp}

\begin{proof}
    See Appendix~\ref{sec:proof-bd}.
\end{proof}
\noindent 
The TLR statistic is biased, in the sense that its value is minimized away from the truth, i.e. for \(\beta\neq\beta_0\). In the following, we show that we can construct a near-unbiased test by recentering the TLR statistic.

\paragraph*{A recentered TLR statistic.}
In order to recenter the TLR statistic, it is necessary to first
consider the signed root of the TLR statistic. The signed-root TLR statistic can be used to test the one-sided hypothesis, \(\beta \leq \beta_0\). We record its definition and limiting distribution as a lemma.

\begin{lemmap}{SLR}[Signed-root TLR Statistic]\label{lem:slr}
Denote the signed root of the TLR statistic,
\begin{align*}
    {\mathrm{SLR}}(\beta_0) \;\equiv\; \operatorname{sign}\!\left(\sum_{j=1}^{2d_Z}\hat\kappa_j\,\hat Q_j^2\right)\sqrt{\mathrm{TLR}(\beta_0)}.
\end{align*}
For \(\sqrt{n}\tau_n=\tau\), as \(n\to\infty\), the limiting distribution is given as,
\begin{align*}
    {\mathrm{SLR}}(\beta_0) \convd \tfrac{1}{\sqrt{2\,d_Z}}\left(\sqrt{(1+\varrho)\,S_+}\,-\,\sqrt{(1-\varrho)\,S_-}\right) .
\end{align*}
Denote this limiting distribution \( T(\varrho,\xi)\).
The  distribution  has non-zero mean for \(\varrho\neq 0\).

\end{lemmap}

\begin{proof}
    See Appendix~\ref{sec:proof-slr}.
\end{proof}
\noindent
The bias of the TLR statistic is associated with the  non-zero mean of its signed root.
We can reduce this bias by subtracting an estimate of this mean. In order to take into account finite-sample deviations from exact eigenvalue homogeneity, we use a parametric bootstrap estimator.
We record the definition of the recentered TLR statistic.

\begin{definitionp}{RTLR}[Recentered TSLS Likelihood-Ratio Statistic]\label{prop:rtlr}
Let \(\hat q^*\) denote the constrained minimizer of the TLR statistic, \(\tilde Q^b \sim\mathfrak{N}(\hat q^*,\,\boldsymbol{\iota}_{2d_Z})\) a parametric bootstrap draw from its limiting distribution, where \(\boldsymbol{\iota}_{2d_Z}\) denotes the identity matrix, and let the bootstrapped TLR statistic be given as,
\begin{align*}
    \mathrm{TLR}^b(\beta_0) = d_Z^{-1}\sum_{j=1}^{2d_Z} (\tilde Q_j^b)^2\!\left(\frac{\tilde\lambda^*\hat\kappa_j}{1+\tilde\lambda^*\hat\kappa_j}\right)^{\!2}
    \quad\text{s.t.}\quad
    \sum_{j=1}^{2d_Z}\frac{\hat\kappa_j (\tilde Q_j^b)^2}{(1+\tilde\lambda^*\hat\kappa_j)^2} = 0.
\end{align*}
Let \(\hat b(\hat q^*, \hat \kappa)\) denote parametric bootstrap estimators of the asymptotic mean of the signed-root TLR statistic given the vector of eigenvalues, \(\hat\kappa\).
Define the recentered TSLS likelihood-ratio statistic as
\begin{align*}
    \mathrm{RTLR}(\beta_0) \equiv {({\mathrm{SLR}}(\beta_0) - \hat b(\hat q^*, \hat \kappa))^2}.
\end{align*}
\end{definitionp}

\noindent
Recentering does not obtain a pivotal statistic. However, 
simulations suggest that
 it is approximately centered across large parts of the parameter space.
As we show in the following, it is possible to leverage both the raw and recentered TLR statistic to
construct  uniformly valid tests for the TSLS hypothesis.

\paragraph*{Robust TLR Inference.}

A test that is valid in both the strong- and weak-instrument regimes must be valid for all realizations of the nuisance parameters, \((\varrho,\xi)\). 
In the following, we introduce a two-step procedure that uses the fact that we can consistently estimate \(\varrho\) under the null, and that \(\hat S\) carries information about \(\xi\), to produce uniformly valid inference, in the spirit of \cite{bergerValuesMaximizedConfidence1994}.

The statistic \(\hat S\) is such that \(d_Z\hat S\) is distributed asymptotically non-central chi-squared with non-centrality parameter  \(\xi\). It follows that we can 
use \(\hat S\) to produce a level \((1-\alpha_1)\) confidence interval for \(\xi\). We can then use the worst-case \((1-\alpha_2)^{\mathrm{th}}\) quantile in this interval as critical value for the test. If \(\alpha_1+\alpha_2 = \alpha\), this two-step procedure is a valid test for \(\beta=\beta_0\) at level \(\alpha\).
We record the result.

\begin{propositionp}{TS}[Two-step Inference]\label{prop:ts}
Let \(\hat \Xi_{1-\alpha_1}\) denote a \(1-\alpha_1\) confidence interval for \(\xi\) obtained by inverting \(\hat S\), and let \(\alpha_2 \equiv \alpha -\alpha_1 \). Let \(\hat T(\beta_0)\) denote the raw, signed-root or recentered TLR statistic, and let
\(c_{1-\alpha_2}(\varrho,\xi) \) denote the \((1-\alpha_2)^{th}\) {quantile} of its limiting distribution.
  We then have:
\begin{align*}
    \limsup_{n\to\infty}\,\sup_{\varrho,\xi}\,\P\!\left(\hat T(\beta_0) > \max_{\xi \in \hat\Xi_{1-\alpha_1}} c_{1-\alpha_2}(\hat\varrho,\xi)\right) \;\leq\; \alpha_1 + \alpha_2 \;=\; \alpha.
\end{align*}
It follows that \(\hat T(\beta_0) > \max_{\xi \in \hat\Xi_{1-\alpha_1}} c_{1-\alpha_2}(\hat\varrho,\xi)\) is a {uniformly} valid test for \(\beta=\beta_0\).
\end{propositionp}
\begin{proof}
    See Appendix~\ref{sec:proof-ts}.
\end{proof}
\noindent
When the first-stage F-statistic is sufficiently small, tests based on the TLR statistic fail to reject any \(\beta_0\) far from the origin. It follows that associated confidence intervals have infinite length.
We record the result as a lemma.
\begin{lemmap}{UC}[Unbounded Confidence Interval]\label{lem:uc}
Let \(\hat F \equiv \lVert (\hat\Sigma_{\hat\gamma}/n)^{-1/2}\hat\gamma\rVert^2/d_Z\) denote the Wald test statistic for the hypothesis that \(\gamma_n=0\).
The two-step TLR test fails to reject \(\beta_0\) for all sufficiently large \(\lvert\beta_0\rvert\) if and only if
\[
    d_Z\hat F \;\leq\; {\chi^2_{d_Z,1-\alpha_2}}.
\]
Equivalently, confidence intervals formed by test inversion have infinite length.
\end{lemmap}
\begin{proof}
    See Appendix~\ref{sec:proof-uc}.
\end{proof}
\noindent
It is instructive to understand how much  the limiting distribution of the TLR and recentered TLR statistics differ for different values of \((\varrho,\xi)\) and \(d_Z\). We study this in Monte Carlo simulations. As the limiting distribution of the TLR statistic is symmetric in \(\varrho\), it suffices to consider positive \(\varrho\).

\begin{figure}[!t]
    \centering

    \includegraphics[width=\textwidth]{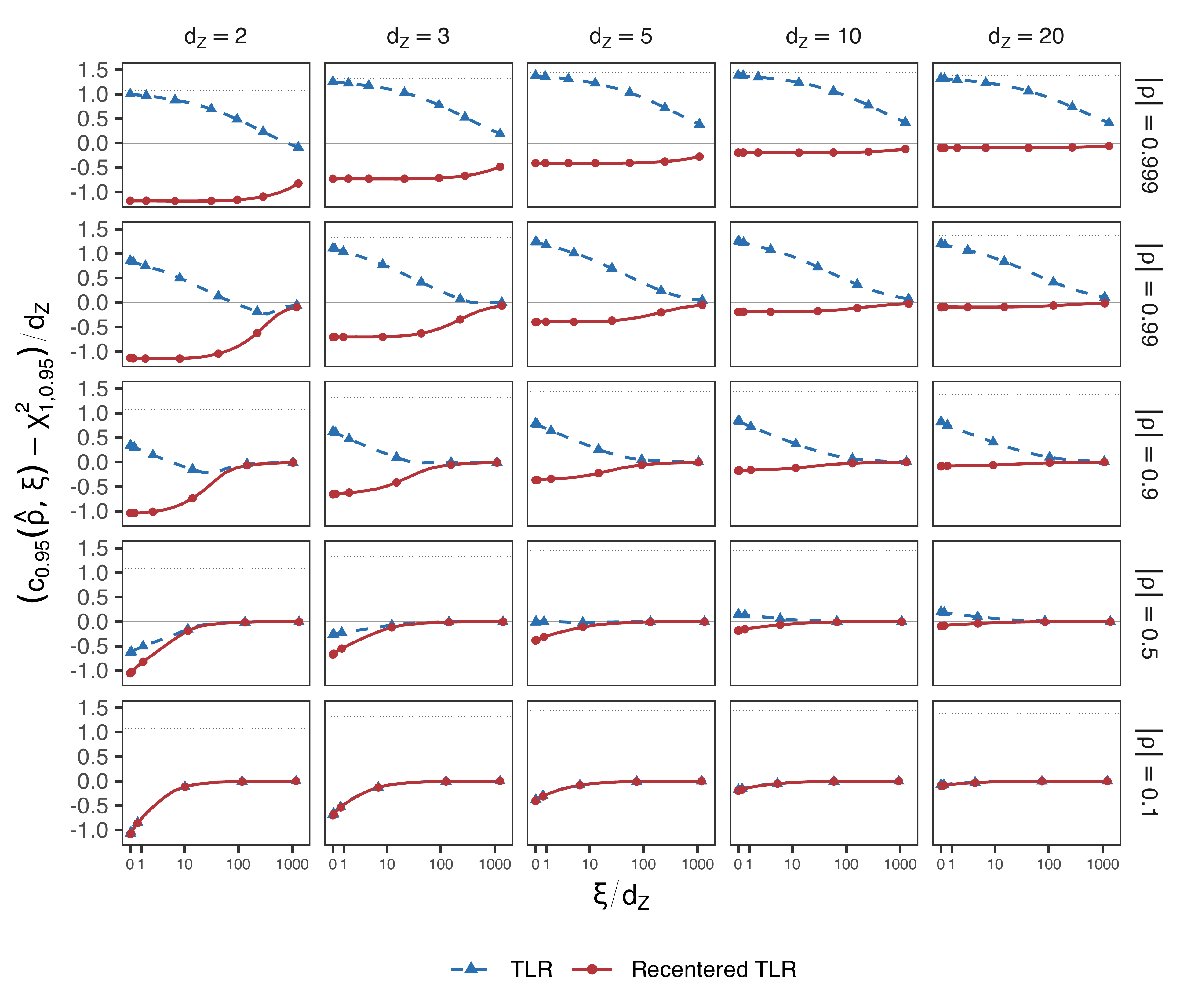}

    \caption{\textbf{Quantiles of the TLR and RTLR Statistics.} This figure plots the \(95^{\mathrm{th}}\) quantile of the TLR and RTLR statistics for different \((\lvert\varrho\rvert,\xi)\) and \(d_Z\). The quantiles are plotted relative to the \(95^{\mathrm{th}}\) quantile of the chi-squared distribution with one degree of freedom and divided by \(d_Z\). The blue line plots the quantiles of the TLR statistic and the red line plots the quantiles of the recentered TLR (RTLR) statistic.  The dashed line indicates \(\chi^2_{d_Z,1-\alpha_2}\), i.e. the limit as \(\lvert\varrho\rvert\to 1\). The x axis  is logarithmic.
    }

    \label{fig:cutoffs-oracle}
\end{figure}

In \Cref{fig:cutoffs-oracle}, we plot the \(95^{\mathrm{th}}\) quantile of the distribution of the TLR statistic and the recentered TLR statistic for different values of \((\varrho, \xi)\) and \(d_Z\). 
We make two observations. First, the \(95^{\mathrm{th}}\) quantile of the TLR statistic meaningfully overshoots the \(95^{\mathrm{th}}\) quantile of the chi-squared distribution with one degree of freedom for a wide range of values of \((\varrho,\xi)\). Second, the \(95^{\mathrm{th}}\) quantile of the recentered TLR statistic is consistently located either at or below the  \(95^{\mathrm{th}}\) quantile of the chi-squared distribution with one degree of freedom. 

From this, we draw two conclusions. First, the relative dominance of the  \(95^{\mathrm{th}}\) quantile of the chi-squared distribution  over that of the recentered TLR statistic suggest that
the combination of the former used as critical value in a test using the recentered TLR statistic will produce an approximately  uniformly valid test, albeit conservative for some values of \((\varrho,\xi)\), and small \(d_Z\). Second, a two-step test with a small \(\alpha_1\approx 0\)
and little if any change in critical values will produce an exactly uniformly valid test.
Given the large-\(\xi\) distribution of the TLR statistic, such a test will numerically recover the Wald test in the strong-instrument limit.
By computing conditional critical values, we can increase the power of this test in the weak-instrument regime.

The two-step test from \Cref{prop:ts} requires the choice of a first-step level \(\alpha_1\). In order to achieve the strong-instrument Wald limit, it is preferable to choose this to be small. To this end, we let \(\alpha_1=10^{-5}\). However, the performance of the test is relatively invariant to this choice in simulations, and one can equivalently choose e.g. \(\alpha_1=10^{-10}\), giving slightly different second-step conditional critical values.

\begin{figure}[!b]
    \centering

    \includegraphics[width=\textwidth]{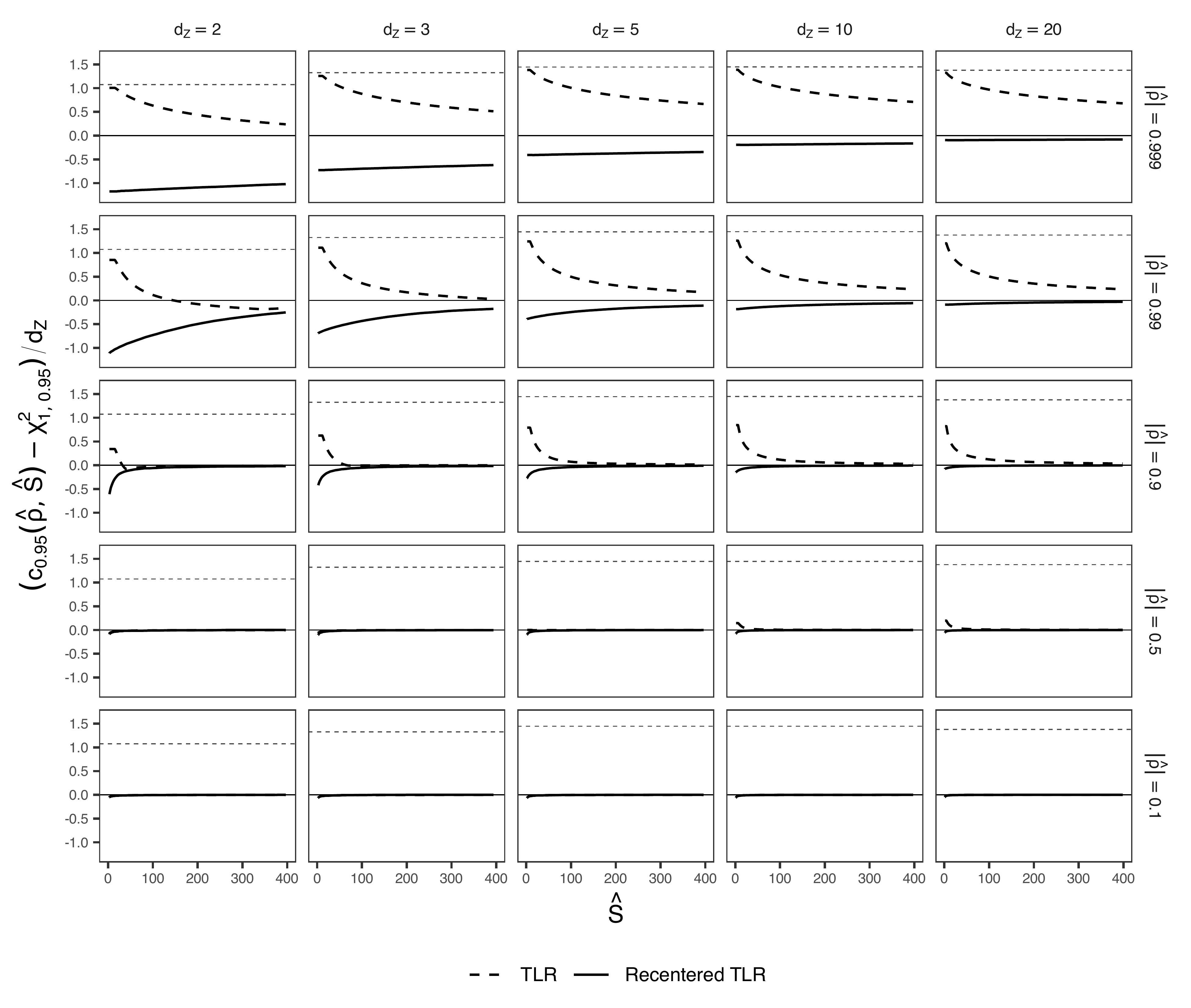}

    \caption{\textbf{Conditional Second-Step Critical Values of the TLR and Recentered TLR Statistics.} This figure plots conditional second-step critical values for an \(\alpha=0.05\) test, with first-step level \(\alpha_1=10^{-5}\), for different values of \((\hat \varrho,\hat S)\) and \(d_Z\).
    The critical values are plotted relative to the \(95^{\mathrm{th}}\) quantile of the chi-squared distribution with one degree of freedom and divided by \(d_Z\). The solid line plots critical values for the recentered TLR statistic and the dashed line plots the quantiles of the TLR statistic.  The thin dashed horizontal line indicates \(\chi^2_{d_Z,1-\alpha_2}\), i.e. the limit as \(\lvert\varrho\rvert\to 1\). The x axis starts at the smallest \(\hat S\) consistent with  \(\hat F > \chi^2_{d_Z,0.95}/d_Z\).
    }

    \label{fig:cutoffs-conditional}
\end{figure}

\Cref{fig:cutoffs-conditional} plots conditional critical values for a level \(\alpha=0.05\) two-step TLR test (dashed) and two-step recentered TLR  test (solid), 
for different values of \((\hat \varrho, \hat S)\) and \(d_Z\). We see that conditional critical values for the
 TLR test differ considerably from its strong-instrument limit.
The conditional critical values of the recentered TLR test are either at or below the \(95^{\mathrm{th}}\) quantile of the chi-squared distribution with one degree of freedom for most values of  \(\hat \varrho\) and \(\hat S\). In practice, such conditional critical values can either be tabulated, as done for the ``VtF'' test introduced by \cite{leeWhatWhenYou2023}, or computed on the fly.

Lastly, it is of interest to understand the power properties of the two-step TLR and recentered tests relative to the Wald test constructed using the TSLS estimator. While this comparison is not prima facie reasonable, as the TSLS Wald test is not valid with weak instruments, it is of interest to understand the extent to which  the two-step TLR and recentered TLR  tests replicate the power properties of the TSLS Wald test with strong and moderately strong instruments.

\begin{figure}[!b]
    \centering
    \includegraphics[width=\textwidth]{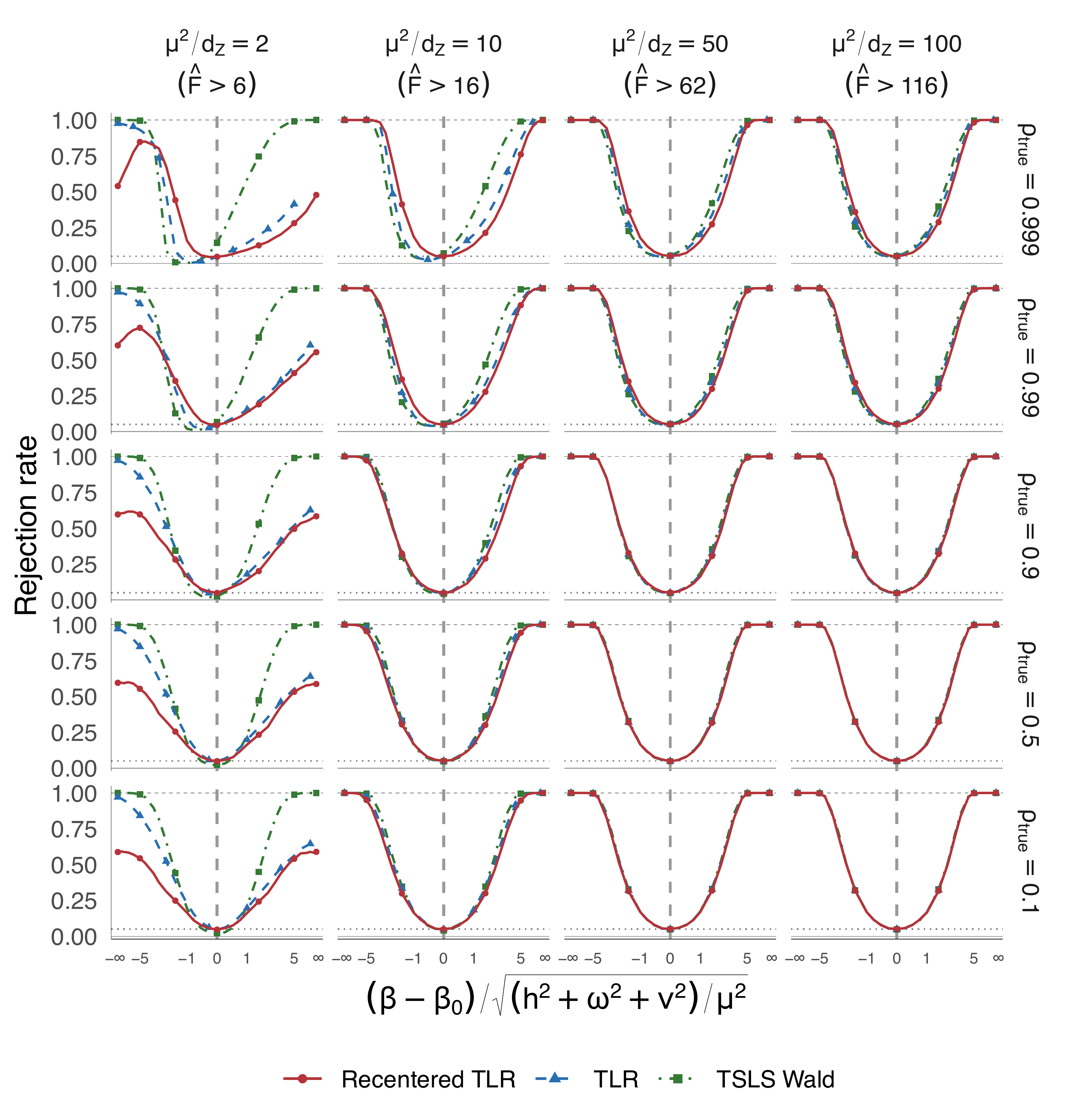}

    \caption{\textbf{Power of the TLR and Recentered TLR Tests.} This figure shows simulations of the power of the two-step TLR and two-step recentered TLR tests against alternatives with \(d_Z=5\) instruments and different degrees of endogeneity. The red line shows the two-step recentered TLR test, the blue line shows the two-step TLR test and the green line shows the TSLS Wald test.
    The y axis shows the rejection rate, and the x axis enumerates distance from the true parameter, in terms of a natural scaling of nuisance parameters. The simulation is conducted with \(\varrho_{\mathrm{true}}\equiv \omega/\sqrt{h^2+\omega^2} \in\{0.1,0.5,0.8,0.99,0.999\}\), displayed across vertical panels.
    }
    \label{fig:lr-regular-asy-all-power}
\end{figure}

In \Cref{fig:lr-regular-asy-all-power} we plot the power of the  TLR and recentered TLR two-step tests in simulations, 
and compare them to the power of the TSLS Wald test, for different values of the concentration parameter, \(\mu^2\), and \(\omega\), fixing \(\nu^2=1\), \(h=0.1\) and \(d_Z=5\). We let  \(\varrho_{\mathrm{true}}=\omega/\sqrt{h^2+\omega^2}\) denote the true value of \(\varrho\) off the null.
For values of the concentration parameter, \(\mu^2\), we indicate the value of the first-stage F-statistic that would reject the hypothesis \(\mu^2\leq\bar \mu^2\) at level \(\alpha=0.05\) in the panel header. Recall from \Cref{lem:pd} that \(\xi\) is a deterministic function of \(\omega,h^2,\nu^2\) and \(\mu^2\), and hence pinned down by these parameters. The x-axis is scaled relative to the value of the nuisance parameters, in order for curves to take comparable shapes across panels. 

Considering \Cref{fig:lr-regular-asy-all-power}, three  observations become apparent. First, the power curve of the recentered TLR statistic is close to being symmetric, and the test is close to being unbiased, for most of the parameter space. The exception is for extreme realizations of  \(\varrho_{\mathrm{true}}\) and \(\xi\). Second, we see that unlike the TSLS Wald test, both the TLR and recentered TLR tests reject no more than 5\% of the time at the null, \(\beta=\beta_0\). The TSLS Wald test, on the other hand, overrejects under the null at small \(\xi\) and large \(\varrho\). Third, we see that as the instrument set grows stronger, the power curves of the three tests converge to the same limit. This suggests that the power of the TLR and recentered TLR tests asymptote to the power of the Wald test.

\section{Conclusion}
\label{sec:conclusion}

In this paper we have shown that existing weak-instrument robust tests are invalid if used as tests for the TSLS estimand in a model with treatment effect heterogeneity and more than one instrument. In particular, we have shown that the AR, Kleibergen--Moreira LM and Moreira CLR tests can overreject at any level of instrument strength, and that the degree of overrejection is increasing as the instrument set grows stronger.

We have further shown that there exists a test statistic that can be leveraged to produce  valid inference with treatment effect heterogeneity. In particular, we have constructed the TSLS likelihood-ratio (TLR) statistic for hypotheses on the TSLS estimand, and derived its limiting distribution. We have shown that when combined  with a pretest in a two-step procedure, the test obtains uniform validity. Choosing a small first-step level, we have shown in simulations that the test inherits the properties of the Wald test when the instrument set is strong.

\clearpage

\appendix
\renewcommand\thesection{\Alph{section}}
\renewcommand\thesubsubsection{\Alph{section}.\arabic{subsection}.\roman{subsubsection}}

\renewcommand\thesection{\Alph{section}}
\section{Proofs}
\label{app:proofs}

\subsection{Proof of \Cref{lem:st}}
\label{sec:proof-st}
\begin{proof}
By \(\sqrt n\,(\hat\tau-\tau_n)\;\convd\;\mathfrak{N}(\mathbf{0}_{2d_Z},\Sigma)\), \(\hat \Sigma \convp \Sigma\) and
 \(\sqrt n\,\tau_n=\tau\) constant, we have
\[
(\hat\Sigma/n)^{-1/2}\hat\tau \;\convd\; \mathfrak{N}\!\bigl(\Sigma^{-1/2}\tau,\ \boldsymbol{\iota}_{2d_Z}\bigr).
\]
 By \(\xi\equiv \lVert(\Sigma/n)^{-1/2}\tau_n\rVert^2=\lVert\Sigma^{-1/2}\tau\rVert^2\), we thus have \(d_Z\hat S = \lVert(\hat\Sigma/n)^{-1/2}\hat\tau\rVert^2 \convd \mathfrak{C}^2_{2d_Z}(\xi)\).
 
\end{proof}

\subsection{Proof of \Cref{lem:pd}}
\label{sec:proof-pd}
\begin{proof}
Recall that we can write \(\Sigma=\bigl[\begin{smallmatrix}\Sigma_{\hat\delta} & \Sigma_{\hat\delta,\hat\gamma}\\ \Sigma_{\hat\delta,\hat\gamma}' & \Sigma_{\hat\gamma}\end{smallmatrix}\bigr]\).
\paragraph*{Decomposition.}

Let \(B\equiv\Sigma_{\hat\delta,\hat\gamma}\Sigma_{\hat\gamma}^{-1}\) denote the matrix of regression coefficients in the joint regression of the components of the vector \(\hat\delta\) on the components of the vector \(\hat \gamma\). \(\Sigma\) can then be decomposed as follows,
\begin{align}
\Sigma = \begin{bmatrix} \boldsymbol{\iota}_{d_Z} & B \\ \mathbf{0}_{d_Z\times d_Z} & \boldsymbol{\iota}_{d_Z} \end{bmatrix}
\begin{bmatrix} \Sigma_{\hat\delta\mid\hat\gamma} & \mathbf{0}_{d_Z\times d_Z} \\ \mathbf{0}_{d_Z\times d_Z} & \Sigma_{\hat\gamma} \end{bmatrix}
\begin{bmatrix} \boldsymbol{\iota}_{d_Z} & \mathbf{0}_{d_Z\times d_Z} \\ B' & \boldsymbol{\iota}_{d_Z} \end{bmatrix},
\end{align}
where $\Sigma_{\hat\delta\mid\hat\gamma} \equiv \Sigma_{\hat\delta} - B\,\Sigma_{\hat\gamma}\,B' = \Sigma_{\hat\delta}
- \Sigma_{\hat\delta,\hat\gamma}\Sigma_{\hat\gamma}^{-1}\Sigma_{\hat\delta,\hat\gamma}'$. It follows that we can write the inverse of \(\Sigma\) as,
\begin{align*}
\Sigma^{-1} = \begin{bmatrix} \boldsymbol{\iota}_{d_Z} & \mathbf{0}_{d_Z\times d_Z} \\ -B' & \boldsymbol{\iota}_{d_Z} \end{bmatrix}
\begin{bmatrix} \Sigma_{\hat\delta\mid\hat\gamma}^{-1} & \mathbf{0}_{d_Z\times d_Z} \\ \mathbf{0}_{d_Z\times d_Z} & \Sigma_{\hat\gamma}^{-1} \end{bmatrix}
\begin{bmatrix} \boldsymbol{\iota}_{d_Z} & -B \\ \mathbf{0}_{d_Z\times d_Z} & \boldsymbol{\iota}_{d_Z} \end{bmatrix}
= \begin{bmatrix}\Sigma_{\hat\delta\mid\hat\gamma}^{-1} & -\Sigma_{\hat\delta\mid\hat\gamma}^{-1} B\\ -B'\Sigma_{\hat\delta\mid\hat\gamma}^{-1} & \Sigma_{\hat\gamma}^{-1}+B'\Sigma_{\hat\delta\mid\hat\gamma}^{-1}B\end{bmatrix}.
\end{align*}
Applying this decomposition to \(\xi = n\,\tau_n'\Sigma^{-1}\tau_n\), we have:
\begin{align}
\xi 
= n\,(\delta_n-B\gamma_n)'\Sigma_{\hat\delta\mid\hat\gamma}^{-1}(\delta_n-B\gamma_n)
+ \underbrace{n\,\gamma_n'\Sigma_{\hat\gamma}^{-1}\gamma_n}_{\equiv \mu^2}.
\label{eq:pd-schur}
\end{align}
In order to simplify the first term, it will be useful to show that \(\Sigma\) admits an asymptotic Kronecker factorization along the weak-instrument sequence.

\paragraph*{Asymptotic Kronecker form.}
The blocks of \(\Sigma\) generally take the sandwich form,
\begin{align*}
\Sigma_{\hat\delta} =\Sigma_{ZZ}^{-1}\E[ Z_i Z_i' W_i^2]\,\Sigma_{ZZ}^{-1},\
\Sigma_{\hat\gamma} =\Sigma_{ZZ}^{-1}\E[ Z_i Z_i' V_i^2]\,\Sigma_{ZZ}^{-1},\
\Sigma_{\hat\delta,\hat\gamma} =\Sigma_{ZZ}^{-1}\E[ Z_i Z_i' W_iV_i]\,\Sigma_{ZZ}^{-1}.
\end{align*}
As \(n\to\infty\) with \(\sqrt n\,\tau_n=\tau\) fixed, \(\Sigma\) admits a Kronecker form. In particular, we have
\begin{align*}
\E\left[\begin{matrix}
W_i^2 & W_iV_i \\
W_iV_i & V_i^2 \\ 
\end{matrix} \ \middle|  \ Z_i=z\right]
= 
\begin{bmatrix}
\var[W_i] & \cov[W_i,V_i] \\
\cov[W_i,V_i] & \var[V_i] \\ 
\end{bmatrix} +
\begin{bmatrix}
o(1) &o(1) \\
o(1) &o(1) \\
\end{bmatrix}
\end{align*}
such that
\begin{align}
    \Sigma \,\conv\, \begin{bmatrix}\var[W_i] & \cov[W_i,V_i]\\ \cov[W_i,V_i] & \var[V_i]\end{bmatrix}\otimes\Sigma_{ZZ}^{-1}.
\label{eq:pd-kron}
\end{align}
We prove this in the following. In terms of the model we have the regression residuals as,
\begin{align*}
    W_i \equiv \beta_iD_i(Z_i) + U_i - \delta_n'Z_i  
\qquad \qquad 
V_i \equiv D_i(Z_i)  - \gamma_n'Z_i
\end{align*}
Conditional on \(Z_i=z\), the linear terms \(\delta_n'z,\gamma_n'z\)  are constant, and enter conditional moments only through a remainder of order
\(\lVert z\rVert\,\lVert(\delta_n,\gamma_n)\rVert=O(\lVert z\rVert/\sqrt n)\) or \(O(\lVert z\rVert^2/n)\) when squared.
By \Cref{ass:ivx}, \((\beta_i,U_i,\{D_i(z)\}_{z\in\R^{d_Z}})\indep Z_i\), so for any integrable functional
\(g\), we have
\(
\E[g(\beta_i,U_i,D_i(z))\mid Z_i=z]=\E[g(\beta_i,U_i,D_i(z))]
\).
It follows that,
\begin{align}
\begin{aligned}
\E[W_i^2\mid Z_i=z]   &= \E[(\beta_iD_i(z)+U_i)^2]+O(\lVert z\rVert/\sqrt n),\\
\E[W_iV_i\mid Z_i=z]  &= \E[(\beta_iD_i(z)+U_i)D_i(z)]+O(\lVert z\rVert/\sqrt n),\\
\E[V_i^2\mid Z_i=z]   &= \E[D_i(z)^2]+O(\lVert z\rVert/\sqrt n).
\end{aligned}
\label{eq:three-equations}
\end{align}
It thus suffices to show that the leading terms are asymptotically constant in \(z\). 
Observe  first that by Cauchy-Schwarz, we have uniformly in \(z,z'\) for any \(A_i\) with \(\E[A_i^2]<\infty\), that
\begin{align}
\begin{aligned}
    \left|\E[A_iD_i(z)]-\E[A_iD_i(z')]\right|
&\le \E[A_i^2]^{1/2}\,\E\bigl[(D_i(z)-D_i(z'))^2\bigr]^{1/2} \\ 
&\qquad\le \E[A_i^2]^{1/2}\left[\sup_{z,z'}\E\bigl[(D_i(z)-D_i(z'))^2\bigr]\right]^{1/2}
=o(1)
\label{eq:cs-linear}
\end{aligned}
\end{align}
which follows from  \Cref{ass:ivw}, where we have used that \(\E[(D_i(z)-D_i(z'))^2]=\P[D_i(z)\neq D_i(z')]\)  in the binary case.
Second, we have uniformly,
\begin{align}
\begin{aligned}
&\left|\E[A_iD_i(z)^2]-\E[A_iD_i(z')^2]\right| \\ 
&\qquad\le \left[\sup_{z,z'}\E\bigl[(D_i(z)-D_i(z'))^2\bigr]\right]^{1/2}\,\bigl(2\E[A_i^2D_i(z)^2]+2\E[A_i^2D_i(z')^2]\bigr)^{1/2}=o(1)
\end{aligned}
\label{eq:cs-quad}
\end{align}
where we have used that we can write \(D_i(z)^2-D_i(z')^2=(D_i(z)-D_i(z'))(D_i(z)+D_i(z'))\), and that  \(\sup_z\E[A_i^2D_i(z)^2]<\infty\).
Applying
the inequalities from equations \eqref{eq:cs-linear}--\eqref{eq:cs-quad} to \(A_i\in\{1,\beta_i,\beta_i^2,U_i,\beta_iU_i\}\) we see that \(\E[D_i(z)^2]\),
\(\E[\beta_iD_i(z)^2]+\E[U_iD_i(z)]\) and
\(\E[\beta_i^2D_i(z)^2]+2\E[\beta_iU_iD_i(z)]+\E[U_i^2]\) are asymptotically constant in \(z\).
Let \(E_i\in\{W_i^2,V_i^2,W_iV_i\}\) and \(E_i(z)\in\{(\beta_iD_i(z)+U_i)^2,D_i(z)^2,(\beta_iD_i(z) + U_i)D_i(z)\}\). By \cref{eq:three-equations}, we have
\begin{align}
    \E[E_i \mid Z_i=z] = \E[E_i(z)] + r_n(z)
\end{align}
with \(r_n(z)\leq C\lVert z\rVert /\sqrt{n}\) for some constant \(C\), and  \(\E[E_i(z)]=\E[E_i]+o(1)\) uniformly in \(z\) by the law of iterated expectations.
It follows that
\begin{align*}
    &\lVert\E[Z_iZ_i'E_i] -\E[E_i]\Sigma_{ZZ} \rVert \\ 
    &\qquad=   \lVert\E[Z_iZ_i'(\E[E_i(Z_i)] - \E[E_i])] + \E[Z_iZ_i'r_n(Z_i)]\rVert \\
    &\qquad\leq   \lVert\E[Z_iZ_i'(\E[E_i(Z_i)] - \E[E_i])]\rVert + \lVert \E[Z_iZ_i'r_n(Z_i)]\rVert \\
    &\qquad\leq   \underbrace{\sup_{z}\lvert\E[E_i(z)] - \E[E_i]\rvert}_{=o(1)}\ \underbrace{\E[\lVert Z_i\rVert^2]}_{<\infty} + \underbrace{(C/\sqrt{n})}_{=o(1)}\ \underbrace{\E[\lVert Z_i \rVert^3]}_{<\infty} = o(1)
\end{align*}
and thus
\begin{align}
    \Sigma \to \begin{bmatrix}\var[W_i] & \cov[W_i,V_i]\\ \cov[W_i,V_i] & \var[V_i]\end{bmatrix}\otimes\Sigma_{ZZ}^{-1}.
\end{align}
\paragraph*{Simplifying the first term.}

By the Kronecker form on \(\Sigma\), we have 
\begin{align}
    B \conv \frac{\cov[W_i,V_i]}{\var[V_i]} \boldsymbol{\iota}_{d_Z} = \beta_{\mathrm{ols}}\boldsymbol{\iota}_{d_Z}
\end{align}
where \(\boldsymbol{\iota}_{d_Z}\) denotes the \(d_Z \times d_Z\) identity matrix. We further have \(\Sigma_{\hat\gamma}\to\var[V_i]\,\Sigma_{ZZ}^{-1}\) and
\begin{align}
    \Sigma_{\hat\delta\mid\hat\gamma}\conv\Bigl(\var[W_i]-\frac{\cov[W_i,V_i]^2}{\var[V_i]}\Bigr)\,\Sigma_{ZZ}^{-1} \;=\; h^2\,\Sigma_{\hat\gamma},
\end{align}
where  \(h^2\equiv(\var[W_i]-\beta_{\mathrm{ols}}^2\var[V_i])/\var[V_i]\). It further follows that \( \Sigma_{\hat\delta\mid\hat\gamma}^{-1}\conv h^{-2}\Sigma_{\hat\gamma}^{-1}\). 
Substituting \(\delta_n=\beta\gamma_n+(\delta_n-\beta\gamma_n)\) and using that \(B\to\beta_{\mathrm{ols}}\boldsymbol{\iota}\), we thus obtain:
\begin{align*}
\delta_n - B\gamma_n \;=\; (\delta_n-\beta\gamma_n)\,-\,\omega\,\gamma_n + o_P(\lVert\tau_n\rVert),
\end{align*}
where \( \omega\equiv\beta_{\mathrm{ols}}-\beta\).
Inserting for the first term of \cref{eq:pd-schur}, we thus obtain
\begin{align*}
&n(\delta_n-B\gamma_n)'\Sigma_{\hat\delta\mid\hat\gamma}^{-1}(\delta_n-B\gamma_n)\\
&\quad= h^{-2}\bigl[\,n(\delta_n-\beta\gamma_n)'\Sigma_{\hat\gamma}^{-1}(\delta_n-\beta\gamma_n)-\; 2\omega\,n\gamma_n'\Sigma_{\hat\gamma}^{-1}(\delta_n-\beta\gamma_n)\;+\;\omega^2\,n\gamma_n'\Sigma_{\hat\gamma}^{-1}\gamma_n\,\bigr] + o(1)
\end{align*}
Since \(\sqrt n\gamma_n=\gamma\) and \(\sqrt n\delta_n=\delta\), \(\sqrt n(\delta_n-\beta\gamma_n)=\delta-\beta\gamma\), and \(\Sigma_{\hat\gamma}^{-1}\to\var[V_i]^{-1}\Sigma_{ZZ}\), the cross-term becomes
\[n\gamma_n'\Sigma_{\hat\gamma}^{-1}(\delta_n-\beta\gamma_n) \;=\; (\sqrt n\gamma_n)'\Sigma_{\hat\gamma}^{-1}(\sqrt n(\delta_n-\beta\gamma_n)) \;\to\; \var[V_i]^{-1}\,\gamma'\Sigma_{ZZ}(\delta-\beta\gamma).\]
By \(\beta=(\gamma'\Sigma_{ZZ}\gamma)^{-1}\gamma'\Sigma_{ZZ}\delta\), the inner product \(\gamma'\Sigma_{ZZ}(\delta-\beta\gamma)\) equals zero, thus it is \(o(1)\). We obtain,
\begin{align}
    &n(\delta_n-B\gamma_n)'\Sigma_{\hat\delta\mid\hat\gamma}^{-1}(\delta_n-B\gamma_n)\;=\;\mu^2 h^{-2}(\nu^2 + \omega^2) + o(1),
\end{align}
where \(\nu^2\equiv \lVert\Sigma_{ZZ}^{1/2}(\delta_n-\beta\gamma_n)\rVert^2/\lVert\Sigma_{ZZ}^{1/2}\gamma_n\rVert^2\) and \(\mu^2\equiv \lVert(\Sigma_{\hat \gamma}/n)^{-1/2}\gamma_n\rVert^2\).
Combining,
\[
\xi \;=\; \mu^2 \,+\, \mu^2 h^{-2}(\nu^2 + \omega^2) \;=\; \mu^2\left(1 + h^{-2}(\nu^2+\omega^2)\right),
\]
which was what we set out to show.

\end{proof}

\subsection{Proof of \Cref{lem:eh}}
\label{sec:proof-eh}
\begin{proof}
We prove the claims step by step.

\paragraph*{Eigenvalue signature.}
Observe that the constraint matrix \(\Gamma(\beta_0)\) factors as
\begin{align*}
\Gamma(\beta_0)
=\begin{bmatrix}
    0 & 1\\ 
    1 & -2\beta_0\end{bmatrix}\otimes\Sigma_{ZZ}
= \bigl(\boldsymbol{\iota}_2\otimes\Sigma_{ZZ}^{1/2}\bigr)\,
\left(\begin{bmatrix}
    0 & 1\\ 
    1 & -2\beta_0\end{bmatrix}\otimes\boldsymbol{\iota}_{d_Z}\right)\,
\bigl(\boldsymbol{\iota}_2\otimes\Sigma_{ZZ}^{1/2}\bigr),
\end{align*}
where \(\otimes\) denotes the Kronecker product.
It follows that we can write \(\Lambda(\beta_0)\) as
\begin{align*}
\Lambda(\beta_0)
={\Sigma^{1/2}\bigl(\boldsymbol{\iota}_2\otimes\Sigma_{ZZ}^{1/2}\bigr)}
\,\left(\begin{bmatrix}
    0 & 1\\ 
    1 & -2\beta_0\end{bmatrix}\otimes\boldsymbol{\iota}_{d_Z}\right)\,
{\bigl(\boldsymbol{\iota}_2\otimes\Sigma_{ZZ}^{1/2}\bigr)\Sigma^{1/2}}
\end{align*}
By \(\Sigma\) and \(\Sigma_{ZZ}\) positive definite, the matrix \(\Sigma^{1/2}(\boldsymbol{\iota}_2\otimes\Sigma_{ZZ}^{1/2})\) is nonsingular.
It follows that \(\Lambda(\beta_0)\) is congruent to the matrix \(\left[\begin{smallmatrix}
    0 & 1\\ 
    1 & -2\beta_0\end{smallmatrix}\right]\otimes\boldsymbol{\iota}_{d_Z}\), and by Sylvester's law of inertia the two share signature, i.e. the same number of positive, negative and zero eigenvalues. 

    The eigenvalues of \(\left[\begin{smallmatrix}
    0 & 1\\ 
    1 & -2\beta_0\end{smallmatrix}\right]\) are \(-\beta_0\pm\sqrt{\beta_0^2+1}\): One strictly positive and one strictly negative. It follows that \(\left[\begin{smallmatrix}
    0 & 1\\ 
    1 & -2\beta_0\end{smallmatrix}\right]\otimes\boldsymbol{\iota}_{d_Z}\) has each of these two eigenvalues with multiplicity \(d_Z\), giving eigenvalue signature \((d_Z,d_Z,0)\). It follows that \(\Lambda(\beta_0)\) has \(d_Z\) strictly positive and \(d_Z\) strictly negative eigenvalues. 

\paragraph*{Eigenvalue homogeneity.}
From the proof of \Cref{lem:pd}, \(\Sigma\) has, asymptotically, a Kronecker representation along the weak-instrument asymptotic sequence. Thus, under Assumptions \ref{ass:ivx} and \ref{ass:ivw} \(\Lambda(\beta_0)\) simplifies further. In particular, we have,
\begin{align}
    \Lambda(\beta_0)\;\to\; 
    \left(\begin{bmatrix}\var[W_i] & \cov[W_i,V_i]\\ \cov[W_i,V_i] & \var[V_i]\end{bmatrix}^{\frac{1}{2}}\begin{bmatrix}0&1\\1&-2\beta_0\end{bmatrix}\begin{bmatrix}\var[W_i] & \cov[W_i,V_i]\\ \cov[W_i,V_i] & \var[V_i]\end{bmatrix}^{\frac{1}{2}}\right)\otimes\boldsymbol{\iota}_{d_Z}.
\end{align}
As a consequence, along the weak-instrument asymptotic sequence, the eigenvalues of \(\Lambda(\beta_0)\) are identical to those of the matrix
\(\left[\begin{smallmatrix}\var[W_i] & \cov[W_i,V_i]\\ \cov[W_i,V_i] & \var[V_i]\end{smallmatrix}\right]^{1/2}\left[\begin{smallmatrix}0&1\\1&-2\beta_0\end{smallmatrix}\right]\left[\begin{smallmatrix}\var[W_i] & \cov[W_i,V_i]\\ \cov[W_i,V_i] & \var[V_i]\end{smallmatrix}\right]^{1/2}\), each with multiplicity \(d_Z\). Denote these two eigenvalues \(\kappa_+>0\) and \(\kappa_-<0\).
By continuity of the spectrum of symmetric matrices, it follows that for all \(k=1,\dots,d_Z\), \(\kappa^{(k)}_+\to\kappa_+\) and \(\kappa^{(k)}_-\to\kappa_-\).
 
\paragraph*{Constraint in canonical coordinates.}
By definition, we have \(q\equiv X'\Sigma^{-1/2}\sqrt{n}\,\tau_n\). The orthogonal matrix \(X\) diagonalises \(\Lambda(\beta_0)\), with eigenvalues \(\kappa_j\). We thus have,
\begin{align*}
    n\,\tau_n'\Gamma(\beta_0)\tau_n
    =\sqrt{n}\,\tau_n'\Sigma^{-1/2}\Lambda(\beta_0)\Sigma^{-1/2}\sqrt{n}\tau_n
    = q'\diag(\kappa)\,q
    =\sum_{j:\kappa_j>0}\kappa_j q_j^2 + \sum_{j:\kappa_j<0}\kappa_j q_j^2.
\end{align*}
By \(\kappa_j\to\{\kappa_+,\kappa_-\}\), we have
\(\kappa_j\to\kappa_+\) on the positive eigenspace and \(\kappa_j\to\kappa_-\) on the negative eigenspace, and the constraint \(\tau_n'\Gamma(\beta_0)\tau_n=0\) is equivalent to
\begin{align*}
    \kappa_+\lVert q_+\rVert^2 \;=\; \lvert\kappa_-\rvert\lVert q_-\rVert^2.
\end{align*}
By definition, \(\varrho=(\kappa_+-|\kappa_-|)/(\kappa_++|\kappa_-|)\). Solving for \(\kappa_+\) and \(\kappa_-\), we obtain \(\kappa_+=\tfrac{1}{2}(1+\varrho)(\kappa_++\lvert\kappa_-\rvert)\) and \(\lvert\kappa_-\rvert=\tfrac{1}{2}(1-\varrho)(\kappa_++\lvert\kappa_-\rvert)\). Inserting for these and  dividing  by \(\tfrac{1}{2}(\kappa_++\lvert\kappa_-\rvert)\) yields the constraint, 
\[(1+\varrho)\lVert q_+\rVert^2=(1-\varrho)\lVert q_-\rVert^2.\]

\paragraph*{Closed-form expression for \(\varrho\).}

The eigenvalues of the matrix \(\Lambda(\beta_0)\) are given by the eigenvalues of the matrix
\begin{align}
    \tilde\Lambda(\beta_0) \equiv\begin{bmatrix}\var[W_i] & \cov[W_i,V_i]\\ \cov[W_i,V_i] & \var[V_i]\end{bmatrix}\begin{bmatrix}0&1\\1&-2\beta_0\end{bmatrix}.
\end{align}
We thus have:
\begin{align*}
    \kappa_+ + \kappa_- &= \tr[\tilde\Lambda(\beta_0)] = 2\bigl(\cov[W_i,V_i] - \beta_0\var[V_i]\bigr) \\ 
    \kappa_+ \cdot \kappa_-&= \det[\tilde\Lambda(\beta_0)] =  -\bigl(\var[W_i]\,\var[V_i] - \cov[W_i,V_i]^2\bigr).
\end{align*}
The discriminant of the characteristic polynomial is
\begin{align*}
(\kappa_+ - \kappa_-)^2 &= (\kappa_++\kappa_-)^2 - 4\kappa_+\kappa_-
= 4\,\var[V_i]\,\var[W_i-\beta_0V_i],
\end{align*}
Observe that \(\var[W_i-\beta_0V_i]=\var[W_i]-2\beta_0\cov[W_i,V_i]+\beta_0^2\var[V_i]\). Hence
\begin{align}
\varrho \;=\; \frac{\kappa_+-\lvert\kappa_-\rvert}{\kappa_++\lvert\kappa_-\rvert} \;=\; \frac{\kappa_+ +\kappa_-}{\kappa_+-\kappa_-} \;=\; \frac{\cov[W_i,V_i]-\beta_0\var[V_i]}{\sqrt{\var[V_i]\,\var[W_i-\beta_0V_i]}}.
\end{align}
Recall from the proof of \Cref{lem:pd} that we have
\(\cov[W_i,V_i]\conv\beta_{\mathrm{ols}}\var[V_i]\) and \(\var[W_i]\conv\var[V_i](h^2+\beta_{\mathrm{ols}}^2)\).
Letting \(\omega_0\equiv\beta_{\mathrm{ols}}-\beta_0\), the numerator is \(\omega_0\var[V_i]\) and
\begin{align*}
\var[W_i-\beta_0V_i] \;&\conv\; \var[V_i](h^2+\beta_{\mathrm{ols}}^2) - 2\beta_0\beta_{\mathrm{ols}}\var[V_i] + \beta_0^2\var[V_i]
= \var[V_i]\,(h^2 + \omega_0^2).
\end{align*}
It follows that
\(\sqrt{\var[V_i]\,\var[W_i-\beta_0V_i]} = \var[V_i]\sqrt{h^2+\omega_0^2}\), and we have,
\[
\varrho = \frac{\omega_0\,\var[V_i]}{\var[V_i]\,\sqrt{h^2+\omega_0^2}}
\]
and hence \(\varrho = \omega_0/\sqrt{h^2+\omega_0^2}\)
\end{proof}

\subsection{Proof of \Cref{lem:rob-cte}}
\label{sec:proof-rob-cte}
\begin{proof}
Observe that \(\hat\Omega(\beta_0)\equiv \hat\Sigma_{\hat\delta} - \beta_0\bigl(\hat\Sigma_{\hat\delta,\hat\gamma}+\hat\Sigma_{\hat\delta,\hat\gamma}'\bigr) + \beta_0^2\,\hat\Sigma_{\hat\gamma}\),
\[
\tilde\gamma(\beta_0) \;\equiv\; \hat\gamma \;-\; \bigl(\hat\Sigma_{\hat\delta,\hat\gamma}' - \beta_0\,\hat\Sigma_{\hat\gamma}\bigr)\,\hat\Omega(\beta_0)^{-1}\,\bigl(\hat\delta - \beta_0\hat\gamma\bigr)
\]
and
\(
\hat\Psi(\beta_0)\;\equiv\;\hat\Sigma_{\hat\gamma}-\bigl(\hat\Sigma_{\hat\delta,\hat\gamma}'-\beta_0\hat\Sigma_{\hat\gamma}\bigr)\hat\Omega(\beta_0)^{-1}\bigl(\hat\Sigma_{\hat\delta,\hat\gamma}-\beta_0\hat\Sigma_{\hat\gamma}\bigr)
\).
For simplicity of exposition, define
\[
\Delta \;\equiv\; (\hat\Omega(\beta_0)/n)^{-1/2}(\hat\delta-\beta_0\hat\gamma),\qquad
\Pi \;\equiv\; (\hat\Omega(\beta_0)/n)^{-1/2}\tilde\gamma(\beta_0).
\]
Under the composite null, \(\delta_n=\beta_0\gamma_n\), we have \(\sqrt n(\hat\delta-\beta_0\hat\gamma)\convd\mathfrak{N}(0,\Omega(\beta_0))\). It follows that
\(\Delta\convd\mathfrak{N}(0,\boldsymbol{\iota}_{d_Z})\).
{Applying Slutsky to replace \(\hat\Omega\), \(\hat\Sigma\) by their probability limits,  \(\sqrt n(\hat\delta-\beta_0\hat\gamma)\) and \(\sqrt n\,\tilde\gamma(\beta_0)\) are linear functionals of \(\sqrt n(\hat\tau-\tau_n)\), hence are distributed jointly asymptotically Normal. The asymptotic covariance of \(\sqrt n\,\hat\gamma\) with \(\sqrt n(\hat\delta-\beta_0\hat\gamma)\) is \(\Sigma_{\hat\delta,\hat\gamma}'-\beta_0\Sigma_{\hat\gamma}\). Subtracting \((\Sigma_{\hat\delta,\hat\gamma}'-\beta_0\Sigma_{\hat\gamma})\Omega(\beta_0)^{-1}(\hat\delta-\beta_0\hat\gamma)\) to generate \(\tilde\gamma\) we obtain  \(\cov[\sqrt n\,\tilde\gamma,\sqrt n(\hat\delta-\beta_0\hat\gamma)]\to 0\).}
By joint asymptotic normality, thus, \(\Pi\indep\Delta\) in the limit. We now turn to proving each of the three limiting results.
\begin{enumerate}
    \item (AR) 
    \(d_Z\mathrm{AR}(\beta_0)=\lVert\Delta\rVert^2\convd\mathfrak{C}^2_{d_Z}\).
    \item (LM)
    The  distribution of \(\Pi'\Delta / \lVert \Pi\rVert\)  conditional on \(\Pi = \pi\) is given as  \(\pi'\Delta/\lVert\pi\rVert\sim\mathfrak{N}(0,1)\). This follows since \(\E[\Delta]=0\) and \(\var[\Delta]=\boldsymbol{\iota}_{d_Z}\).
    The conditional distribution is invariant to \(\pi\). It follows that unconditionally \(\bigl(\Pi'\Delta/\lVert\Pi\rVert\bigr)^2\sim\mathfrak{C}^2_1\), independently of \(\Pi\). It follows that \(d_Z\mathrm{LM}(\beta_0)\equiv(\Pi'\Delta)^2/\lVert\Pi\rVert^2\convd\mathfrak{C}^2_1\).
    \item  
     (CLR)
      Observe that we can write \(\mathrm{AR}(\beta_0)=\mathrm{LM}(\beta_0)+\mathrm{J}(\beta_0)\) where \(\mathrm{J}(\beta_0)\equiv\mathrm{AR}(\beta_0)-\mathrm{LM}(\beta_0)\), and  \(\mathrm{J}(\beta_0)\) is the squared norm of the projection of \(\Delta\) onto the orthogonal complement of \(\Pi/\lVert\Pi\rVert\).
        Conditional \(\Pi/\lVert\Pi\rVert\), the parallel and orthogonal projections of  \(\Delta\sim\mathfrak{N}(0,\boldsymbol{\iota})\) are independent, by rotational invariance.
        It follows that, unconditionally, \(\mathrm{J}(\beta_0)\indep\mathrm{LM}(\beta_0)\) with \(d_Z\mathrm{J}(\beta_0)\convd\mathfrak{C}^2_{d_Z-1}\).
        The CLR statistic is a continuous function of \(d_Z\mathrm{AR}(\beta_0)\), \(d_Z\mathrm{LM}(\beta_0)\), and \(\lVert\hat r(\beta_0)\rVert^2\). Applying the continuous mapping theorem with \(\mathrm{AR}(\beta_0)\stackrel{d}{=}\mathrm{J}(\beta_0)+\mathrm{LM}(\beta_0)\) in the limit, with \(\hat r(\beta_0)\) a function of \(\Pi\) only and independent of \(\Delta\), gives the result.
\end{enumerate}

\end{proof}

\subsection{Proof of \Cref{lem:rob-hte}}
\label{sec:proof-rob-hte}
\begin{proof}
Let \(\Delta\) and \(\Pi\) be defined as in the proof of \Cref{lem:rob-cte}.
Under the TSLS null \(\beta=\beta_0\), given the asymptotic Kronecker structure of \(\Sigma\) from the proof of \Cref{lem:pd}, and the fact that \(\cov[W_i,V_i]\conv\beta_{\mathrm{ols}}\,\var[V_i]\) and \(\var[W_i]\conv\var[V_i](h^2+\beta_{\mathrm{ols}}^2)\), we have
\begin{align}\label{eq:leg-hte-omega}
\Omega(\beta_0) \;&=\; \var[V_i]\,(h^2+\omega^2)\,\Sigma_{ZZ}^{-1}+o(1)
\end{align}
and \(\Omega(\beta_0)^{-1}(\Sigma_{\hat\delta,\hat\gamma}'-\beta_0\Sigma_{\hat\gamma}) \to\frac{\omega}{h^2+\omega^2}\boldsymbol{\iota}_{d_Z}\).
By  definition of \(\nu^2\), we further have,
\begin{align}
    n(\delta_n-\beta_0\gamma_n)'\Sigma_{ZZ}(\delta_n-\beta_0\gamma_n) = \var[V_i]\,\mu^2\nu^2 + o(1)
\end{align}
Further,  by \(\mu^2 \equiv n\gamma_n'\Sigma_{\hat\gamma}^{-1}\gamma_n\) and \(\Sigma_{\hat\gamma}\to\var[V_i]\,\Sigma_{ZZ}^{-1}\), we have
\(n\gamma_n'\Sigma_{ZZ}\gamma_n = \var[V_i]\,\mu^2 + o(1)\).
By the moment condition from the TSLS estimand, we have \(n\gamma_n'\Sigma_{ZZ}(\delta_n-\beta_0\gamma_n)=0\). We prove each of the three statements, and then the size distortion result.

\paragraph*{AR.}
We have 
\(\sqrt n(\hat\delta-\beta_0\hat\gamma)\convd\mathfrak{N}(\delta-\beta_0\gamma,\Omega(\beta_0))\). It follows that 
\begin{align}
\Delta\convd\mathfrak{N}(\Omega(\beta_0)^{-1/2}(\delta-\beta_0\gamma),\boldsymbol{\iota}_{d_Z}).
\end{align}
Let \(\theta_{\mathrm{AR}}\equiv\lVert \Omega(\beta_0)^{-1/2}(\delta-\beta_0\gamma)\rVert^2\).
We then have:
\begin{align}
\theta_{\mathrm{AR}}= \frac{n(\delta_n-\beta_0\gamma_n)'\Sigma_{ZZ}(\delta_n-\beta_0\gamma_n)}{\var[V_i](h^2+\omega^2)}
= \frac{\mu^2\nu^2}{h^2+\omega^2}.
\end{align}
It follows that
\(d_Z\mathrm{AR}(\beta_0)=\lVert\Delta\rVert^2\convd\mathfrak{C}^2_{d_Z}(\theta_{\mathrm{AR}})\).

\paragraph*{LM.}
By \cref{eq:leg-hte-omega}, along the weak-instrument asymptotic sequence, we have
\begin{align}
\sqrt n\,\tilde\gamma(\beta_0) \;=\; \sqrt n\,\hat\gamma - \tfrac{\omega}{h^2+\omega^2}\sqrt n(\hat\delta-\beta_0\hat\gamma)+o_P(1).
\end{align}
where \(o_P(1)\) follows by Slutsky, as \((\hat\Sigma_{\hat\delta,\hat\gamma}'-\beta_0\hat\Sigma_{\hat\gamma})\hat\Omega(\beta_0)^{-1}\convp\frac{\omega}{h^2+\omega^2}\boldsymbol{\iota}_{d_Z}\).
It follows that
\begin{align}
\Pi\convd\mathfrak{N}\left(\Omega(\beta_0)^{-1/2}\bigl(\gamma-\tfrac{\omega}{h^2+\omega^2}(\delta-\beta_0\gamma)\bigr),\tfrac{h^2}{(h^2+\omega^2)^2}\,\boldsymbol{\iota}_{d_Z}\right)
\end{align}
where the variance is derived by observing that \(\var[\sqrt n\,\tilde\gamma]\to\Sigma_{\hat\gamma}-2\tfrac{\omega}{h^2+\omega^2}(\Sigma_{\hat\delta,\hat\gamma}'-\beta_0\Sigma_{\hat\gamma})+\tfrac{\omega^2}{(h^2+\omega^2)^2}\Omega(\beta_0)\to\tfrac{h^2}{h^2+\omega^2}\var[V_i]\,\Sigma_{ZZ}^{-1}\), and
 \(\Pi\indep\Delta\) by the same argument as in the proof of \Cref{lem:rob-cte}. By adding and subtracting, we can decompose the LM statistic into a standard normal and a non-central distortion,
\begin{align}\label{eq:hte-lm-decomp}
\sqrt{d_Z\mathrm{LM}(\beta_0)} 
= \underbrace{\frac{\Pi'(\Omega(\beta_0)^{-1/2}(\delta-\beta_0\gamma))}{\lVert\Pi\rVert}}_{\convd\,\Upsilon} \;+\; \underbrace{\frac{\Pi'(\Delta - \Omega(\beta_0)^{-1/2}(\delta-\beta_0\gamma))}{\lVert\Pi\rVert}}_{\convd\,\zeta}.
\end{align}
where \(\zeta\) is equivalent to the root of the constant-effects LM-statistic, and similarly to the proof of \Cref{lem:rob-cte}, we have \(\zeta\sim\mathfrak{N}(0,1)\) conditional on \(\Pi=\pi\).  The conditional distribution is invariant in \(\pi\), hence \(\zeta\sim\mathfrak{N}(0,1)\) also unconditionally, with \(\zeta\indep\Pi\), and thus also \(\zeta\indep\Upsilon\). It remains to derive the expression for \(\Upsilon\).

Consider decomposing \(\Pi\) into the component parallel to and orthogonal to
\(\delta_n-\beta_0\gamma_n\) normalized by the root of the asymptotic variance, and denote the limiting distribution of these two components \(\zeta_1\) and  \(\zeta_{d_Z-1}\). We have
\begin{align*}
\zeta_1
&\equiv
\frac{h^2+\omega^2}{h}\cdot
\frac{\Pi'\,\Omega(\beta_0)^{-1/2}(\delta_n-\beta_0\gamma_n)}
     {\lVert\Omega(\beta_0)^{-1/2}(\delta_n-\beta_0\gamma_n)\rVert},
\\[6pt]
\zeta_{d_Z-1}
&\equiv
\frac{h^2+\omega^2}{h}\,
\cdot\,
\left[\Pi
-\frac{\Pi'\,\Omega(\beta_0)^{-1/2}(\delta_n-\beta_0\gamma_n)}
       {\lVert\Omega(\beta_0)^{-1/2}(\delta_n-\beta_0\gamma_n)\rVert^2}\,
\Omega(\beta_0)^{-1/2}(\delta_n-\beta_0\gamma_n)\right].
\end{align*}
Let \(\zeta_{d_Z-1}^2\equiv \lVert\zeta_{d_Z-1}\rVert^2\). Observe that \[\tfrac{h^2+\omega^2}{h}\Pi \convd \mathfrak{N}\left(\tfrac{h^2+\omega^2}{h}\,\Omega(\beta_0)^{-1/2}(\gamma-\tfrac{\omega}{h^2+\omega^2}(\delta-\beta_0\gamma)),\boldsymbol{\iota}_{d_Z}\right).\]
Observe that \(\zeta_1\) is the projection of \(\tfrac{h^2+\omega^2}{h}\Pi\) onto the line spanned by \(\Omega(\beta_0)^{-1/2}(\delta-\beta_0\gamma)\), and \(\zeta_{d_Z-1}\) is the orthogonal complement. Since the covariance with \(\tfrac{h^2+\omega^2}{h}\Pi\) is the identity, we have
\begin{align*}
\zeta_1^2\sim\mathfrak C^2_1\!\Bigl(\tfrac{\omega^2\mu^2\nu^2}{h^2(h^2+\omega^2)}\Bigr),
\qquad
\zeta_{d_Z-1}^2\sim\mathfrak C^2_{d_Z-1}\!\Bigl(\tfrac{\mu^2(h^2+\omega^2)}{h^2}\Bigr),
\qquad
\zeta_1^2\indep\zeta_{d_Z-1}^2,
\end{align*}
which follows since
\(\Omega(\beta_0)\conv(h^2+\omega^2)\var[V_i]\Sigma_{ZZ}^{-1}\),
\(\lVert\sqrt{n}\Sigma_{ZZ}^{1/2}(\delta_n-\beta_0\gamma_n)\rVert^2=\mu^2\nu^2\var[V_i]\), and
\begin{align}
    \lVert\Omega(\beta_0)^{-1/2}\sqrt n(\delta_n-\beta_0\gamma_n)\rVert^2
&\conv\mu^2\nu^2/(h^2+\omega^2),
\end{align}
under the null.
Further, as \(\zeta\indep\Pi\) and \((\zeta_1,\zeta_{d_Z-1})\) is a function of \(\Pi\), we have \(\zeta\indep(\zeta_1^2,\zeta_{d_Z-1}^2)\). This gives
\begin{align}
\Upsilon^2
=\frac{(\Pi'\Omega(\beta_0)^{-1/2}\sqrt n(\delta_n-\beta_0\gamma_n))^2}{\lVert\Pi\rVert^2\,\lVert\Omega(\beta_0)^{-1/2}\sqrt n(\delta_n-\beta_0\gamma_n)\rVert^2}\cdot\lVert\Omega(\beta_0)^{-1/2}\sqrt n(\delta_n-\beta_0\gamma_n)\rVert^2
\label{eq:hte-upsilon}
\end{align}
and it follows that
\begin{align}
    \Upsilon^2 \convd
    \theta_{\mathrm{AR}}\cdot\frac{\zeta_1^2}{\zeta_1^2+\zeta_{d_Z-1}^2}.
\end{align}
where \(\theta_{\mathrm{AR}}=\frac{\mu^2\nu^2}{h^2+\omega^2}\).

\paragraph*{CLR.}
Recall that  
\( \mathrm{J}(\beta_0)\equiv\mathrm{AR}(\beta_0)-\mathrm{LM}(\beta_0)\) is the squared norm of \(\Delta\) 
projected onto the orthogonal complement of \(\Pi/\lVert\Pi\rVert\). Conditional on \(\Pi=\pi\), the parallel and orthogonal projections are  independent, and 
\( d_Z\,\mathrm{J}(\beta_0)\mid\Pi\sim\mathfrak{C}^2_{d_Z-1}(\theta_{\mathrm{AR}}-\Upsilon^2) = \mathfrak{C}^2_{d_Z}(\theta_{\mathrm{AR}})- (\Upsilon+\zeta)^2\). By the continuous mapping theorem and \(m(\,\cdot\,,\,\cdot\,,\,\cdot\,)\) given in \Cref{lem:rob-cte}, and the observation that
\(\hat r(\beta_0)=(\hat\Psi(\beta_0)/n)^{-1/2}\tilde\gamma(\beta_0)\) is asymptotically a function only of \(\Pi\) and hence independent of \(\Delta\), the result obtains.

\paragraph*{Size distortion.} 
\begin{enumerate}
    \item 
    (AR) Observe that whenever \(\nu^2>0\), we have \(\theta_{\mathrm{AR}}>0\), and hence \(\mathfrak{C}^2_{d_Z}(\theta_{\mathrm{AR}}) \succ_{\mathrm{FOSD}} \mathfrak{C}^2_{d_Z}\). It follows that
    \begin{align}\label{eq:hte-ar-sizedistort}
\limsup_{n\to\infty}\,\P\!\left[\mathrm{AR}(\beta_0)> c^{\mathrm{AR}}_{1-\alpha}\right]
\;=\;\P\!\left[\mathfrak{C}^2_{d_Z}(\theta_{\mathrm{AR}}) \geq \chi^2_{d_Z,1-\alpha}\right]
\;>\;\alpha,
\end{align}
Since \(\lim_{\mu^2\to\infty} \theta_{\mathrm{AR}}=\infty\), we further have
\begin{align}\label{eq:hte-ar-power}
\lim_{\mu^2\to\infty}\limsup_{n\to\infty}\,\P\!\left[\mathrm{AR}(\beta_0)> c^{\mathrm{AR}}_{1-\alpha}\right]
\;=\;\lim_{\mu^2\to\infty}\P\!\left[\mathfrak{C}^2_{d_Z}(\theta_{\mathrm{AR}}) \geq \chi^2_{d_Z,1-\alpha}\right] = 1
\end{align}
    \item 
    (LM)
Assume \(\omega^2>0\). Then the noncentrality of \(\zeta_1\) is
\(\omega^2\mu^2\nu^2/(h^2(h^2+\omega^2))>0\), so
\(\zeta_1^2\sim\mathfrak{C}^2_1(\omega^2\mu^2\nu^2/[h^2(h^2+\omega^2)])
\succ_{\mathrm{FOSD}}\mathfrak{C}^2_1\). It follows that the expectation of  \(\zeta_1^2/(\zeta_1^2+\zeta_{d_Z-1}^2)\) is strictly positive, 
and the mapping \(u\mapsto\P[(u+\zeta)^2\geq\chi^2_{1,1-\alpha}]\)
is strictly convex and symmetric with minimum \(\alpha\) at \(u=0\), and
\begin{align}\label{eq:hte-lm-sizedistort}
\limsup_{n\to\infty}\,\P\!\left[\mathrm{LM}(\beta_0)> c^{\mathrm{LM}}_{1-\alpha}\right]
\;=\;\E_\Upsilon\!\left[\P\!\left[(\Upsilon+\zeta)^2\geq\chi^2_{1,1-\alpha}\mid\Upsilon\right]\right]
\;>\;\alpha,
\end{align}
by Jensen's inequality. 
Further, let \(\theta_{\mathrm{LM},1}\equiv \omega^2\mu^2\nu^2/(h^2(h^2+\omega^2))\) denote the non-centrality parameter of \(\zeta_1^2\) and \(\theta_{\mathrm{LM},d_Z-1}\equiv \mu^2(h^2+\omega^2)/h^2\) the non-centrality parameter of \(\zeta_{d_Z-1}^2\), and observe that
\(\lim_{\mu^2\to\infty}\theta_{\mathrm{LM},1}=\lim_{\mu^2\to\infty}\theta_{\mathrm{LM},d_Z-1}=\infty\), with divergence at the same linear rate. Dividing and multiplying, we have,
\begin{align}
    \frac{\zeta_1^2}{\zeta_1^2+\zeta_{d_Z-1}^2}
    = \frac{(\zeta_1^2/\theta_{\mathrm{LM},1})\theta_{\mathrm{LM},1}}{(\zeta_1^2/\theta_{\mathrm{LM},1})\theta_{\mathrm{LM},1}+(\zeta_{d_Z-1}^2/\theta_{\mathrm{LM},d_Z-1})\theta_{\mathrm{LM},d_Z-1}}
\end{align}
and since \(\mathfrak{C}^2_k(\theta)/\theta\convp1\) as \(\theta\to\infty\), we obtain
\begin{align}
\frac{\zeta_1^2}{\zeta_1^2+\zeta_{d_Z-1}^2}
    \;\convp\;
\frac{\omega^2\nu^2}{(h^2+\omega^2)^2+\omega^2\nu^2}\;\in(0,1),
\end{align}
Since \(\Upsilon^2=\theta_{\mathrm{AR}}\cdot\zeta_1^2/(\zeta_1^2+\zeta_{d_Z-1}^2)\), we thus have that \(\Upsilon^2\) diverges linearly in \(\mu^2\), and 
\begin{align}\label{eq:hte-lm-power}
\lim_{\mu^2\to\infty}\limsup_{n\to\infty}\,\P\!\left[\mathrm{LM}(\beta_0)> c^{\mathrm{LM}}_{1-\alpha}\right]
= \lim_{\mu^2\to\infty}\E_\Upsilon\!\left[\P\!\left[(\Upsilon+\zeta)^2\geq\chi^2_{1,1-\alpha}\mid\Upsilon\right]\right]
\;=\;1.
\end{align}
    \item 
    (CLR)
    Assume \(\omega^2>0\).
    The limit of \(\textrm{CLR}(\beta_0)\) is 
    \(m\bigl((\Upsilon+\zeta)^2,\,\mathfrak{C}^2_{d_Z}(\theta_{\mathrm{AR}})-(\Upsilon+\zeta)^2,\,\lVert\hat r(\beta_0)\rVert^2\bigr)\), increasing in its first two arguments. As shown above, whenever \(\nu^2>0\), we have \(\theta_{\mathrm{AR}}>0\), hence \(
    \mathfrak{C}^2_{d_Z}(\theta_{\mathrm{AR}})-(\Upsilon+\zeta)^2 \succ_{\mathrm{FOSD}} \mathfrak{C}_{d_Z}^2 - \mathfrak{C}_{1}^2 = \mathfrak{C}_{d_Z-1}^2
    \). Further, \((\Upsilon+\zeta)^2 \succeq_{\mathrm{FOSD}} \mathfrak{C}^2_1\) as shown above. By monotonicity of \(m\) it follows that
    \begin{align}\label{eq:hte-clr-sizedistort}
    \limsup_{n\to\infty}\,\P\!\left[\mathrm{CLR}(\beta_0)> c^{\mathrm{CLR}}_{1-\alpha}\right]
    \;>\;\alpha.
    \end{align}
    By (1), \(\mathfrak{C}^2_{d_Z}(\theta_{\mathrm{AR}})\) diverges as \(\mu^2\to\infty\). By (2), so does \((\Upsilon+\zeta)^2\). Lastly, by the same argument as for \(\Pi\), \(\hat r(\beta_0)\) is asymptotically \(\mathfrak{N}(\Psi(\beta_0)^{-1/2}(\gamma-\tfrac{\omega}{h^2+\omega^2}(\delta-\beta_0\gamma)),\,\boldsymbol{\iota}_{d_Z})\) with \(\Psi(\beta_0)=\tfrac{h^2}{h^2+\omega^2}\var[V_i]\,\Sigma_{ZZ}^{-1}\). Using that \(n\gamma_n'\Sigma_{ZZ}(\delta_n-\beta_0\gamma_n)=0\), the squared norm of the mean equals \(\mu^2\bigl[(h^2+\omega^2)^2+\omega^2\nu^2\bigr]/\bigl(h^2(h^2+\omega^2)\bigr)\), so
\begin{align}\label{eq:hte-rhat-limit}
\lVert\hat r(\beta_0)\rVert^2 \;\convd\; \mathfrak{C}^2_{d_Z}\!\left(\frac{\mu^2\bigl[(h^2+\omega^2)^2+\omega^2\nu^2\bigr]}{h^2(h^2+\omega^2)}\right) = O_P(\mu^2).
\end{align}
Since \(m\) is monotone and unbounded in its first two arguments,
     \(m\bigl((\Upsilon+\zeta)^2,\,\mathfrak{C}^2_{d_Z}(\theta_{\mathrm{AR}})-(\Upsilon+\zeta)^2,\,\lVert\hat r(\beta_0)\rVert^2\bigr)\) grows without bound as its first two arguments diverge, and 
     \begin{align}\label{eq:hte-clr-power}
    \lim_{\mu^2\to\infty}\limsup_{n\to\infty}\,\P\!\left[\mathrm{CLR}(\beta_0)> c^{\mathrm{CLR}}_{1-\alpha}\right]
    =1.
    \end{align}
\end{enumerate}

\end{proof}

\subsection{Proof of \Cref{prop:tlr}}
\label{sec:proof-tlr}
\begin{proof}
Using that 
\(\hat Q\equiv\hat X'\hat\Sigma^{-1/2}\sqrt n\,\hat\tau\) and \(q\equiv X'\Sigma^{-1/2}\sqrt n\,\tau_n\), the objective satisfies \(n(\hat\tau-\tau_n)'\hat\Sigma^{-1}(\hat\tau-\tau_n)=\lVert\hat Q-q\rVert^2\), and we can rewrite the quadratic program as,
\begin{align}
    d_Z^{-1}\min_{q\in \R^{2d_Z}}\lVert\hat Q-q\rVert^2 \qquad \text{s.t.} \qquad \sum_{j=1}^{2d_Z}\hat\kappa_j q_j^2=0.
    \label{eq:optimization-prog}
\end{align}
We use this formulation to derive the statistic.
Let
\begin{align}
    \mathcal{L}(q,\lambda) \equiv \sum_j(\hat Q_j-q_j)^2+\lambda\sum_j\hat\kappa_jq_j^2
\end{align}
denote the Lagrangian. The first-order condition for \(q_j\) is 
\begin{align}
    0=\frac{\partial \mathcal L(q,\lambda)}{\partial q_j} = -2(\hat Q_j - q_j) +2\lambda \hat\kappa_j q_j
\end{align}
which solves to \(\hat Q_j  =q_j (1+\lambda\hat\kappa_j )\). Hence the minimizer is
\begin{align}
    q_j^*(\lambda)=\frac{\hat Q_j}{1+\lambda\hat\kappa_j},
    \label{eq:tlr-qopt}
\end{align}
as long as \(1+\lambda\hat\kappa_j\neq 0\) for all \(j\). The Hessian of \(\mathcal L(q,\lambda)\) in \(q\) is \(2(\boldsymbol{\iota}_{2d_Z}+\lambda\diag(\hat\kappa))\). This matrix is positive definite {if and only if} \(1+\lambda\hat\kappa_j>0\) for all \(j\). Consider this condition separately for positive and negative \(\hat\kappa_j\). In the positive case, the condition implies \(\lambda>-1/\hat \kappa_j\), and  binds at \(\max_{j:\hat\kappa_j>0}\hat\kappa_j\). In the negative case, the condition implies \(\lambda<-1/\hat \kappa_j\), and binds at \(\min_{j:\hat\kappa_j<0}\hat\kappa_j\). It follows that the Hessian is positive definite on 
\[
\lambda\in\left(-1/\max_{j:\hat\kappa_j>0}\hat\kappa_j,\,-1/\min_{j:\hat\kappa_j<0}\hat\kappa_j\right) \equiv \mathcal I
\]
which is
an open interval containing \(0\), with oppositely signed  endpoints. Substituting the first-order condition from \cref{eq:tlr-qopt} into the constraint, \(\sum_j\hat\kappa_jq_j^2=0\), we obtain,
\begin{equation}\label{eq:tlr-secular}
\sum_{j=1}^{2d_Z}\frac{\hat\kappa_j\hat Q_j^2}{(1+\lambda\hat\kappa_j)^2}=0.
\end{equation}
Define now \(\psi(\lambda)\equiv \sum_{j=1}^{2d_Z}\frac{\hat\kappa_j\hat Q_j^2}{(1+\lambda\hat\kappa_j)^2}\). 
 \(\psi(\lambda)\) is smooth on \(\mathcal I\), with
 \begin{align}
 \psi'(\lambda)=-2\sum_{j=1}^{2d_Z}\frac{\hat\kappa_j^2\hat Q_j^2}{(1+\lambda\hat\kappa_j)^3}<0    
 \label{eq:derivative-tlr}
 \end{align}
 where \((1+\lambda\hat\kappa_j)^3\) is positive on \(\mathcal I\) and
 \(\hat\kappa_j^2\hat Q_j^2\geq 0\) with at least one strictly positive, by \(\Sigma\) and \(\Sigma_{ZZ}\) positive definite. We further have
 \begin{align}
     \lim_{\lambda \downarrow -1/\max_{j:\hat\kappa_j>0} \hat\kappa_j} \psi(\lambda) = \infty
     \qquad \text{and} \qquad
     \lim_{\lambda \uparrow -1/\min_{j:\hat\kappa_j<0} \hat\kappa_j} \psi(\lambda) = -\infty
 \end{align}
 This obtains by seeing that the denominator approaches \(0\) from above in either case. By the intermediate value theorem and \cref{eq:derivative-tlr}, there is a unique \(\lambda^* \in\mathcal I\) such that \(\psi(\lambda^*)=0\).
By inserting for \(q^*(\lambda^*)\) in  \cref{eq:optimization-prog} we obtain:
\begin{align}
    \mathrm{TLR}(\beta_0)=d_Z^{-1}\lVert\hat Q-q^*(\lambda^*)\rVert^2=d_Z^{-1}\sum_{j=1}^{2d_Z}\hat Q_j^2\!\left(\frac{\lambda^*\hat\kappa_j}{1+\lambda^*\hat\kappa_j}\right)^{\!2},
\end{align}
where \(\lambda^*\) solves~\cref{eq:tlr-secular}. Within \(\mathcal I\), \(q^*(\lambda^*)\) is the unconstrained minimizer of \(\mathcal{L}(q,\lambda^*)\).
By \citet[Theorem 3.2]{sternIndefiniteTrustRegion1995}, this is also the unique global minimizer of the constrained problem.

\end{proof}

\subsection{Proof of \Cref{prop:ld}}
\label{sec:proof-ld}
\begin{proof}

We prove the strong- and weak-instrument result separately.

\paragraph*{Strong-instrument limit.}
Consider first the setting of \(\tau_n\) fixed, \(n\to\infty\).
The TLR statistic is  the asymptotic likelihood-ratio statistic for the restriction \(g(\tau_n,\beta_0)\equiv\tau_n'\hat\Gamma(\beta_0)\tau_n=0\).
If the conditions of Wilks' theorem are fulfilled, this likelihood-ratio statistic is distributed asymptotically chi-squared with one degree of freedom.
We verify the three conditions as given in Theorem~12.4.2(c) of \cite{lehmannTestingStatisticalHypotheses2005}.
\begin{enumerate}
    \item \textbf{Quadratic-mean differentiability with positive-definite Fisher information.} \(\sqrt{n}(\hat \tau-\tau_n)\convd \mathfrak{N}(\mathbf 0_{2d_Z},\Sigma)\), hence asymptotically, \(\hat \tau \sim \mathfrak{N}(\tau_n,\Sigma/n)\), which is a location family, and is quadratic-mean differentiable with Fisher information \(\Sigma^{-1}\). Since \(\Sigma\) is positive definite, we have \(\Sigma^{-1} \succ 0\).
    
    \item \textbf{Restriction is continuously differentiable.} 
   Observe that \(g\) is a polynomial of degree~\(2\) in \(\tau_n\). It follows that \(g\) is infinitely differentiable. The  restriction has dimension \(1\).
    \item \textbf{The Jacobian of the restriction has full rank on the null manifold.} 
    The restriction has dimension \(1\), hence rank \(1\) suffices.
    Let \(\hat\Sigma_{ZZ}\) denote a consistent estimator for \(\Sigma_{ZZ}\). 
    By direct computation, we have
    \[\nabla_{\tau_n} g(\tau_n,\beta_0)=2\hat\Gamma(\beta_0)\tau_n=
    \begin{bmatrix}
        2\hat\Sigma_{ZZ}\gamma_n \\
        2\hat\Sigma_{ZZ}\delta_n-4\beta_0\hat\Sigma_{ZZ}\gamma_n
    \end{bmatrix}\]
    With a strong instrument set, we have \(\gamma_n\to\gamma\neq 0\). Since \(\hat\Sigma_{ZZ}\succ 0\), we have the top block as \(\hat\Sigma_{ZZ}\gamma_n\neq 0\).
    It follows that \(\nabla_{\tau_n} g\neq 0\) on the null manifold.
\end{enumerate}
All three conditions are fulfilled. The unscaled Wilks likelihood-ratio statistic  equals \(d_Z\,\mathrm{TLR}(\beta_0)\). It follows that \(d_Z\,\mathrm{TLR}(\beta_0) \convd \mathfrak{C}^2_1\).

\paragraph*{Weak-instrument limit.}
Consider now  \(\sqrt n\,\tau_n=\tau\) fixed, \(n\to\infty\). 
We have  \(\gamma_n\to 0\) and the strong-instrument limit breaks.
We proceed to derive the limit directly from the limit of the semi-closed form from \Cref{prop:tlr}. This limit is given as 
\begin{align}\label{eq:ld-closed-form}
d_Z\,\mathrm{TLR}(\beta_0)\convd\sum_{j=1}^{2d_Z} Q_j^{2}\!\left(\frac{\lambda^{*}\kappa_j}{1+\lambda^{*}\kappa_j}\right)^{\!2}
\quad \text{s.t.}
\quad
\sum_{j=1}^{2d_Z}\frac{\kappa_j Q_j^{2}}{(1+\lambda^{*}\kappa_j)^{2}}=0.
\end{align}
where \(\hat Q \convd Q \sim \mathfrak{N}(q,\boldsymbol{\iota}_{2d_Z})\), and \(Q_j\) denotes an element of \(Q\).
Let \((Q_+,Q_-)\) denote the subvectors of \(Q\) associated with the positive and negative eigenspace, respectively and \(\lambda^*\) is the maximizer of the population program.
Under \Cref{lem:eh}, and by continuity of the projection  of \(\hat\Lambda(\beta_0)\) onto its positive and negative eigenspace, we have \(\hat S_\pm \equiv \lVert\hat Q_\pm\rVert^2 \convd S_\pm \equiv \lVert Q_\pm\rVert^2\).
Further by \Cref{lem:eh} and the continuous mapping theorem, the equation defining \(\lambda^*\) simplifies to,
\begin{align}\label{eq:ld-secular-eh}
\frac{\kappa_+\,S_+}{(1+\lambda^*\kappa_+)^{2}}\;=\;\frac{\lvert\kappa_-\rvert\,S_-}{(1-\lambda^*\lvert\kappa_-\rvert)^{2}}.
\end{align}
{By the same argument as in the proof of \Cref{prop:tlr}, \(\lambda^*\) is uniquely determined on \(\mathcal I_\infty = (-1/\kappa_+,\,1/\lvert\kappa_-\rvert)\). Let \(\hat\lambda^*\) denote the optimizing multiplier from the sample. By joint convergence of \((\hat\kappa,\hat S_+,\hat S_-) {\convd}\, (\kappa,S_+,S_-)\) and the continuous mapping and implicit function theorems, we have \(\hat\lambda^* {\convd}\, \lambda^*\). The objective converges by the continuous mapping theorem.}
Taking the root on each side and multiplying, we obtain
\begin{align}
{(1-\lambda^*\lvert\kappa_-\rvert)\sqrt{\kappa_+\phantom{|}\,S_+}}{}\;=\;\sqrt{\lvert\kappa_-\rvert\,S_-}(1+\lambda^*\kappa_+).    
\end{align}
Solving for \(\lambda^*\) gives
\begin{align}
 \lambda^*\;=\;\frac{\sqrt{\kappa_+\phantom{|}\,S_+}-\sqrt{\lvert\kappa_-\rvert\,S_-}}{\sqrt{\kappa_+\phantom{|}\,S_+}\lvert\kappa_-\rvert+\sqrt{\lvert\kappa_-\rvert\,S_-}\kappa_+}   
\end{align}
Observe that under \Cref{lem:eh}, \cref{eq:ld-closed-form} simplifies to,
\begin{align}\label{eq:ld-closed-form-simplified}
d_Z\,\mathrm{TLR}(\beta_0) \convd
\left(\frac{\lambda^{*}\kappa_+}{1+\lambda^{*}\kappa_+}\right)^{\!2}\!S_+\;+\;\left(\frac{\lambda^{*}\kappa_-}{1+\lambda^{*}\kappa_-}\right)^{\!2}\!S_-
\end{align}
Inserting for \(\lambda^*\) in \cref{eq:ld-closed-form-simplified}, and simplifying, we obtain,
\begin{align}
d_Z\,\mathrm{TLR}(\beta_0)
\convd 
\frac{\left(\sqrt{\phantom{|}\!\kappa_+\phantom{|}\!S_+}-\sqrt{\lvert\kappa_-\rvert\,S_-}\right)^{\!2}}{\kappa_++\lvert\kappa_-\rvert}.    
\end{align}
From the definition of \(\varrho\), we thus obtain 
\begin{align}
d_Z\,\mathrm{TLR}(\beta_0)\;\convd\;\frac{1}{2}\left(\sqrt{(1+\varrho)\,S_+}-\sqrt{(1-\varrho)\,S_-}\right)^{\!2},    
\end{align}
where that \(S_\pm\sim\mathfrak{C}^2_{d_Z}(\lVert q_\pm \rVert^2 )\), and we have \(\lVert q_+\rVert^{2}+\lVert q_-\rVert^{2}= \lVert q\rVert^2 = \xi\). 

By the constraint from \Cref{lem:eh}, \((1+\varrho)\lVert q_+\rVert^{2}=(1-\varrho)\lVert q_-\rVert^{2}\). Combining these two equations and rearranging, we obtain \(\lVert q_\pm\rVert^{2} ={(1\mp\varrho)}\,\xi/2\).
It follows that \(S_\pm\sim\mathfrak{C}^2_{d_Z}({(1\mp\varrho)}\xi/2)\).
Since \(Q \sim \mathfrak{N}(q,\boldsymbol{\iota}_{2d_Z})\), and \(Q_+\), \(Q_-\) are projections onto the orthogonal positive and negative eigenspaces of \(\Lambda(\beta_0)\), we have \(S_+\indep S_-\). Further, since \(\hat S=\|\hat Q\|^2/d_Z\) with \(\hat X\) orthogonal, \(d_Z\hat S = \|\hat\Sigma^{-1/2}\sqrt n\hat\tau\|^2 = \|\hat Q\|^2 = \|\hat Q_+\|^2+\|\hat Q_-\|^2\). Thus in the limit,  \(S=S_++S_-\).

\end{proof}

\subsection{Proof of \Cref{prop:bd}}
\label{sec:proof-bd}
\begin{proof}

We prove the three statements separately.

\paragraph*{1. Limit as  \(\xi\to\infty\), fixed \(\varrho\in(-1,1)\) \(\implies\) Strong-instrument limit.}

Observe that for any \(A \sim\mathfrak{C}^2_{d_Z}(\lambda) \) with \(\lambda\to\infty\), a delta-method expansion of \(\sqrt{\cdot}\) around \(\lambda\)
gives
\(\sqrt{A}-\sqrt{\lambda}\convd B\sim\mathfrak{N}(0,1)\).
In particular, this follows as
\(\lim_{\lambda\to\infty}\lambda^{-1}\var[A]=\lim_{\lambda\to\infty}\lambda^{-1}2(d_Z+2\lambda)= 4\), so \(C \equiv \sqrt{A}-\sqrt{\lambda}\approx (A-\lambda)/(2\sqrt{\lambda})\) has variance \(\lim_{\lambda\to\infty}\var[C]= 1\).
We apply this to \(S_\pm\), using that by \Cref{prop:ld} we have \(S_\pm\sim\mathfrak{C}^2_{d_Z}\bigl(\tfrac{1\mp\varrho}{2}\,\xi\bigr)\), to obtain:
\begin{align}
    \sqrt{S_\pm} =  \sqrt{(1\mp \varrho)\xi/2} + B + o_P(1)
\end{align}
Observe now that we have:
\begin{align*}
\sqrt{(1+\varrho) S_+} &= \sqrt{1+\varrho}\left(\sqrt{(1- \varrho)\xi/2} + B_+ + o_P(1)\right) =  \sqrt{\frac{1-\varrho^2}{2}\xi} + \sqrt{1+\varrho}\cdot B_+ + o_P(1)\\
\sqrt{(1-\varrho)S_-} &= \sqrt{1-\varrho}\left(\sqrt{(1+ \varrho)\xi/2} + B_- + o_P(1)\right) = \sqrt{\frac{1-\varrho^2}{2}\xi} + \sqrt{1-\varrho}\cdot B_- + o_P(1)
\end{align*}
where \(B_+,B_- \sim \mathfrak{N}(0,1)\), \(B_+ \indep B_-\).
Considering the TLR limit we thus obtain as \(\xi\to\infty\),
\begin{align}
    \frac{1}{2}\left(\sqrt{(1+\varrho)\,S_+}-\sqrt{(1-\varrho)\,S_-}\right)^{\!2}\xrightarrow{\xi\to\infty}
    \frac{1}{2}\left(\sqrt{1+\varrho}\,B_+ -\sqrt{1-\varrho}\,B_-\right)^{\!2}\sim\mathfrak{C}^2_1
\end{align}
hence \(d_Z\,\mathrm{TLR}(\beta_0)\convd \mathfrak{C}^2_1\) as \(\xi\to\infty\). This recovers the strong-instrument limit.

\paragraph*{2. Limit as \(\varrho\to\pm 1\), any \(\xi\geq 0\) Anderson--Rubin Limit.}

Recall that under the null and \Cref{lem:eh}, we have
\begin{align}
    (1+\varrho)\lVert q_+\rVert^{2}=(1-\varrho)\lVert q_-\rVert^{2}
\end{align}
Consider first the limiting case of   \(\varrho=1\), which implies
\((1-\varrho)\lVert q_-\rVert^{2} = 0\) and \(1+\varrho=2\).
This implies \(\lVert q_+\rVert^{2} = 0\), which
 further forces  \(\lVert q_-\rVert^{2}=\xi\). We thus obtain
\(S_+\sim\mathfrak{C}^2_{d_Z}(0)\) and \(S_-\sim\mathfrak{C}^2_{d_Z}(\xi)\). By
\((1+\varrho,1-\varrho)=(2,0)\), we obtain
\begin{align}
\frac{1}{2}\left(\sqrt{(1+\varrho)\,S_+}-\sqrt{(1-\varrho)\,S_-}\right)^{\!2}\;=\;\tfrac{1}{2}\bigl(\sqrt{2\,S_+}-0\bigr)^{\!2}\;=\;S_+\;\sim\;\mathfrak{C}^2_{d_Z}
\end{align}
By symmetry, for \(\varrho=-1\), the same result obtains, with \(S_-\) taking the place of \(S_+\). By continuity in \(\varrho\), the conclusion extends to the limit \(\varrho\to\pm 1\).

\paragraph*{3. Limit as \(\xi=0\) and \(\varrho=0\), irrelevant instrument, no endogeneity.}

At \(\xi=0\) we have \(S_\pm \sim\mathfrak{C}^2_{d_Z}\), and \(\varrho=0\) implies \(1+\varrho=1-\varrho=1\). As such we have,
\begin{align}
    \frac{1}{2}\left(\sqrt{(1+\varrho)\,S_+}-\sqrt{(1-\varrho)\,S_-}\right)^{2} =
    \frac{1}{2}\left(\sqrt{S_+}-\sqrt{S_-}\right)^{2}
\end{align}
where \(S_+,S_-\sim\mathfrak{C}^2_{d_Z}\), \(S_+\indep S_-\), and the result obtains.

\end{proof}

\subsection{Proof of \Cref{lem:slr}}
\label{sec:proof-slr}
\begin{proof}

By  \Cref{prop:ld} and the continuous mapping theorem, we have
\begin{align}
    \sqrt{\mathrm{TLR}(\beta_0)}\;\convd\;\tfrac{1}{\sqrt{2d_Z}}\left|\sqrt{(1+\varrho)\,S_+}-\sqrt{(1-\varrho)\,S_-}\right|
    \label{eq:slr-tlr-limit}
\end{align}

By \Cref{lem:eh}, as \(\sqrt{n}\tau_n=\tau\), \(n\to\infty\), we have
\begin{align}
    \sum_j\kappa_j Q_j^2 = \kappa_+\lVert Q_+\rVert^2 + \kappa_-\lVert Q_-\rVert^2 = \kappa_+ S_+ - \lvert\kappa_-\rvert S_-
\end{align}
Hence, the sign is positive if and only if \(\sqrt{\kappa_+\phantom{|} S_+} > \sqrt{\lvert\kappa_-\rvert S_-}\). By the definition of \(\varrho\), this is equivalent to \(\sqrt{(1+\varrho)S_+} > \sqrt{(1-\varrho)S_-}\). Thus,
\begin{align}
    \operatorname{sign}\left(\sum_j\hat\kappa_j\hat Q_j^2\right)
    \convp 
    \operatorname{sign}\left(\sqrt{(1+\varrho)S_+}-\sqrt{(1-\varrho)S_-}\right)
\end{align}
 It follows that
\begin{align}
    \mathrm{SLR}(\beta_0)\;\convd\;\tfrac{1}{\sqrt{2 d_Z}}\bigl(\sqrt{(1+\varrho)\,S_+}-\sqrt{(1-\varrho)\,S_-}\bigr)\;\equiv\;T(\varrho,\xi).
\end{align}
By linearity of expectation, we have,
\begin{align}
    \E[T(\varrho,\xi)] \;=\; \tfrac{1}{\sqrt{2 d_Z}}\left(\sqrt{1+\varrho}\,\E[\sqrt{S_+}]-\sqrt{1-\varrho}\,\E[\sqrt{S_-}]\right).
    \label{eq:slr-mean}
\end{align}
Observe that the \((1/2)^{\mathrm{th}}\) moment of the noncentral chi-squared distribution with noncentrality parameter \(\theta\) is given as
\begin{align}
    \E\left[\sqrt{S}\right] \;=\; \sqrt{2}\,\frac{\Gamma\!\left(\tfrac{d_Z+1}{2}\right)}{\Gamma\!\left(\tfrac{d_Z}{2}\right)}\,{}_1F_1\!\left(-\tfrac{1}{2};\tfrac{d_Z}{2};-\tfrac{\theta}{2}\right),
    \label{eq:noncentral-chi-mean}
\end{align}
where \({}_1F_1\) is the confluent hypergeometric function (see e.g. \citealp[Ch.~29]{johnsonContinuousUnivariateDistributions1995}). 
Substituting \(\theta = (1\mp\varrho)\xi/2\), we obtain
\begin{align}
    \E[\sqrt{S_\pm}] \;=\; \sqrt{2}\,\frac{\Gamma\!\left(\tfrac{d_Z+1}{2}\right)}{\Gamma\!\left(\tfrac{d_Z}{2}\right)}\,{}_1F_1\!\left(-\tfrac{1}{2};\tfrac{d_Z}{2};-\tfrac{(1\mp\varrho)\xi}{4}\right).
    \label{eq:slr-mean-sqrt}
\end{align}
Thus unless \(\varrho=0\), we have \(\E[T(\varrho,\xi)]\neq 0\).
\end{proof}

\subsection{Proof of \Cref{prop:ts}}
\label{sec:proof-ts}
\begin{proof}

By \Cref{lem:st}, \(d_Z\hat S\convd\mathfrak{C}^2_{2d_Z}(\xi)\). It follows that  \(\hat\Xi_{1-\alpha_1}\) formed by inverting \(\hat S\) satisfies
\begin{equation}
\P\bigl[\xi\in\hat\Xi_{1-\alpha_1}\bigr]\;\conv\;1-\alpha_1.
\label{eq:ts-coverage}
\end{equation}
Recall that \(d_Z\mathrm{TLR}(\beta_0)\convd d_Z\,T(\varrho,\xi)^2\) and let \(c_{1-\alpha_2}(\varrho,\xi)\) denote the \((1-\alpha_2)^{\mathrm{th}}\) quantile of this distribution. \((1\mp\varrho)\xi/2\) and \(\sqrt{1\pm\varrho}\) are jointly continuous in \((\varrho,\xi)\) on \([-1,1]\times\R_+\), hence the CDF of \(d_Z\,T(\varrho,\xi)^2\) is jointly continuous in \((\varrho,\xi)\). The CDF is strictly increasing at any positive quantile, so by the implicit function theorem, \((\varrho,\xi)\mapsto c_{1-\alpha_2}(\varrho,\xi)\) is jointly continuous on \([-1,1]\times\R_+\). By \Cref{prop:bd}, we have \(\lim_{\xi\to\infty}c_{1-\alpha_2}(\varrho,\xi)=\chi^2_{1,1-\alpha_2}\), hence continuity extends to \([-1,1]\times\overline{\R}_+\) where  \(\overline{\R}_+\equiv\R_+\cup\{\infty\}\).

Let \(\hat c\equiv\max_{\xi\in\hat\Xi_{1-\alpha_1}} c_{1-\alpha_2}(\hat\varrho,\xi)\) denote the worst-case \((1-\alpha_2)^{\mathrm{th}}\) quantile over the first-step interval, where \(\alpha_2\equiv\alpha-\alpha_1\). {Observe that \(\hat\Xi_{1-\alpha_1}\) is closed and \(\xi\mapsto c_{1-\alpha_2}(\hat\varrho,\xi)\) is continuous on the closure of \(\R_+\), hence the supremum is attained.}
By definition, for  \(\xi\in\hat\Xi_{1-\alpha_1}\), we have \(\hat c\geq c_{1-\alpha_2}(\hat\varrho,\xi)\). It follows that
\[
\bigl\{d_Z\,\mathrm{TLR}(\beta_0)> \hat c\bigr\}\cap\bigl\{\xi\in\hat\Xi_{1-\alpha_1}\bigr\}\;\subseteq\;\bigl\{d_Z\,\mathrm{TLR}(\beta_0)> c_{1-\alpha_2}(\hat\varrho,\xi)\bigr\}.
\]
It thus further follows that
\begin{align}
\P\bigl(d_Z\,\mathrm{TLR}(\beta_0)> \hat c\bigr)\;\leq\;{\P\bigl[\xi\notin\hat\Xi_{1-\alpha_1}\bigr]}+{\P\bigl[d_Z\,\mathrm{TLR}(\beta_0)> c_{1-\alpha_2}(\hat\varrho,\xi)\bigr]}.
\label{eq:ts-union}
\end{align}
and \(\P[\xi\notin\hat\Xi_{1-\alpha_1}]\to\alpha_1\).
By \(\hat\varrho\convp\varrho\) and continuity of \(\varrho \mapsto c_{1-\alpha_2}(\varrho,\xi)\), we have \(c_{1-\alpha_2}(\hat\varrho,\xi)\convp c_{1-\alpha_2}(\varrho,\xi)\). By Slutsky, we thus have
\[
\P\bigl[d_Z\,\mathrm{TLR}(\beta_0)> c_{1-\alpha_2}(\hat\varrho,\xi)\bigr]\conv\P\bigl(d_Z\,T(\varrho,\xi)^2> c_{1-\alpha_2}(\varrho,\xi)\bigr)=\alpha_2.
\]
Combining, we get
\[
\limsup_{n\to\infty}\,\P\bigl[d_Z\,\mathrm{TLR}(\beta_0)> \hat c\bigr]\leq\alpha_1+\alpha_2\;=\;\alpha,
\]
It follows that the test \(d_Z\,\mathrm{TLR}(\beta_0)> \hat c\) is asymptotically valid at level \(\alpha\) for the hypothesis \(\beta=\beta_0\).
By joint continuity of the mapping \((\varrho,\xi)\mapsto d_Z\,T(\varrho,\xi)^2\) on \([-1,1]\times\overline{\R}_+\), {the probability of rejection is uniformly continuous. Combined with \(\hat\varrho\convp\varrho\) and the uniform coverage of \(\hat\Xi_{1-\alpha_1}\) implied by \Cref{lem:st}, this gives uniform convergence in \((\varrho,\xi)\) of both terms in the union bound from \eqref{eq:ts-union}, we have}
\begin{align*}
\limsup_{n\to\infty}\sup_{(\varrho,\xi)}\,\P\bigl[d_Z\,\mathrm{TLR}(\beta_0)>c_{1-\alpha_2}(\hat\varrho,\xi)\bigr] &\;\leq\;\alpha_2,\\
\limsup_{n\to\infty}\sup_{\xi}\,\P\bigl[\xi\notin\hat\Xi_{1-\alpha_1}\bigr] &\;\leq\;\alpha_1.
\end{align*}
and thus \(\limsup_{n\to\infty}\sup_{(\varrho,\xi)}\P[d_Z\,\mathrm{TLR}(\beta_0)>\hat c]\leq\alpha_1 + \alpha_2 = \alpha\).

{The proof extends to the signed-root statistic,  as \(\mathrm{SLR}(\beta_0)\convd T(\varrho,\xi)\). The CDF of \(T(\varrho,\xi)\) is jointly continuous in \((\varrho,\xi)\) on \([-1,1]\times\overline{\R}_+\), hence the union bound, Slutsky, and uniform continuity goes through with \(d_Z\,\mathrm{TLR}(\beta_0)\) replaced by \(\mathrm{SLR}(\beta_0)\) and \(d_Z\,T(\varrho,\xi)^2\) by \(T(\varrho,\xi)\).}

{The proof extends to the recentered TLR, as \(\hat\kappa\convp\kappa\) and \(\hat q^*\convd q^*\) jointly, where \(q^*\) is the limit of the constrained minimizer of \(\lVert Q-\,q\,\rVert^2\), for
  \(Q\sim\mathfrak{N}(q,\boldsymbol{\iota}_{2d_Z})\) under the null, and
  \(\mathrm{SLR}(\beta_0)\convd T(\varrho,\xi)\). The parametric bootstrap mean, \(\hat b(\hat q^*,\hat\kappa)\), is jointly continuous
   in \((\hat q^*,\hat\kappa)\). Hence, by the continuous mapping theorem, we have \(\mathrm{RTLR}(\beta_0)\convd (T(\varrho,\xi)-b(q^*,\kappa))^2\). The limit distribution is jointly
  continuous in \((\varrho,\xi)\) on \([-1,1]\times\overline{\R}_+\). It follows that the  union bound, Slutsky, and uniform continuity  goes through.}

\end{proof}

\subsection{Proof of \Cref{lem:uc}}
\label{sec:proof-uc}
\begin{proof}

Denote \(\varrho=\varrho(\beta_0)\) to make dependence on \(\beta=\beta_0\) explicit in this proof.
Observe that by \Cref{lem:eh}, and \(\omega_0 \equiv \beta_{\mathrm{ols}} - \beta_0\), we have
\[
\varrho(\beta_0)\;=\;\frac{\beta_{\mathrm{ols}}-\beta_0}{\sqrt{(\beta_{\mathrm{ols}}-\beta_0)^{2}+h^{2}}}.
\]
Fix  \((\beta_{\mathrm{ols}},h)\) and consider
\(\beta_0\to\pm\infty\). We have \(\lim_{\beta_0\to \pm \infty}\varrho(\beta_0) = \mp 1\).
We proceed to show that \(\lim_{\lvert\beta_0\rvert\to\infty}d_Z\mathrm{TLR}(\beta_0) = d_Z \hat F\).

\paragraph*{Bounding \(d_Z\mathrm{TLR}(\beta_0)\) from above.}
{Let \(\hat\Sigma_{ZZ}\) denote a consistent estimator for \(\Sigma_{ZZ}\).}
By \Cref{prop:tlr}, writing out \(\tau_n= \left[\begin{smallmatrix}
    \delta_n \\
    \gamma_n
\end{smallmatrix}\right]\) and \(\hat\tau= \left[\begin{smallmatrix}
    \hat\delta \\
    \hat\gamma
\end{smallmatrix}\right]\) the TLR statistic is defined in terms of the program
  \begin{align}\label{eq:tlr-program}
  \begin{aligned}
  &d_Z\,\mathrm{TLR}(\beta_0)\;=\;\min_{(\delta_n,\gamma_n)\in\R^{2d_Z}}\;n\bigl((\hat\delta,\hat\gamma)-(\delta_n,\gamma_n)\bigr)'\hat\Sigma^{-1}\bigl((\hat\delta,\hat\gamma)-(\delta_n,\gamma_n)\bigr)
  \\ 
  &\qquad\qquad\qquad\qquad\qquad\qquad\text{s.t.}\quad \gamma_n'\hat\Sigma_{ZZ}(\delta_n-\beta_0\gamma_n)=0.
  \end{aligned}
  \end{align}
    Observe that we have for any
  \((x,w),(y,z)\in\R^{d_Z}\times\R^{d_Z}\),
  \begin{align}\label{eq:schur-marg}
  \min_{x\in\R^{d_Z}}\,(x,z)'\hat\Sigma^{-1}(x,z) \;=\; z'\hat\Sigma_{\hat\gamma}^{-1}z,
  \qquad
  \min_{w\in\R^{d_Z}}\,(y,w)'\hat\Sigma^{-1}(y,w) \;=\; y'\hat\Sigma_{\hat\delta}^{-1}y.
  \end{align}
  which can be seen by completing the square in \(x\) and \(w\).
At \(\gamma_n=0\), the constraint reduces to \(\mathbf{0}_{d_Z}'\hat \Sigma_{ZZ}\delta_n=0\), trivially attained for any \(\delta\). Applying the identity to the objective at  \(\gamma_n=0\), we obtain
\begin{align}
  \min_{\delta_n}\,n\,(\hat\delta-\delta_n,\hat\gamma)'\hat\Sigma^{-1}(\hat\delta-\delta_n,\hat\gamma) \;=\;
  n\,\hat\gamma'\hat\Sigma_{\hat\gamma}^{-1}\hat\gamma \;=\; d_Z\,\hat F.
\end{align}
{Since \(\gamma_n=0\) makes the constraint trivial, the minimum is achieved at a feasible value of \((\delta_n,\gamma_n)\) for every \(\beta_0\), giving}
  \begin{align}\label{eq:tlr-univ-bound}
  d_Z\,\mathrm{TLR}(\beta_0)\;\leq\;d_Z\,\hat F\qquad
  \end{align}
for every \(\beta_0\in\R\).

\paragraph*{Bounding \(d_Z\mathrm{TLR}(\beta_0)\) from below.}
We proceed to show 
 \[\liminf_{\lvert\beta_0\rvert\to\infty}d_Z\,\mathrm{TLR}(\beta_0)\geq d_Z\,\hat F.\] 
 Assume  for contradiction 
 that there exist
  \(\epsilon>0\) and a sequence \(\lvert\beta_0^{(k)}\rvert\to\infty\) such that
  \[
  d_Z\,\mathrm{TLR}(\beta_0^{(k)})\;\leq\; d_Z\,\hat F-\epsilon
  \]
  for every \(k\).
  For each \(k\), let \((\delta_n^{(k)},\gamma_n^{(k)})\) be such that they attain the constrained minimum in \cref{eq:tlr-program} at \(\beta_0^{(k)}\). Applying 
  \cref{eq:schur-marg} to \((\hat\delta-\delta_n^{(k)},\hat\gamma-\gamma_n^{(k)})\), we obtain
   \begin{align}\label{eq:gamma-bound}
   \begin{aligned}
  n\,(\hat\gamma-\gamma_n^{(k)})'\hat\Sigma_{\hat\gamma}^{-1}(\hat\gamma-\gamma_n^{(k)})
  & \leq n\,(\hat\delta-\delta_n^{(k)},\hat\gamma-\gamma_n^{(k)})'\hat\Sigma^{-1}(\hat\delta-\delta_n^{(k)},\hat\gamma-\gamma_n^{(k)}) \\
  &= d_Z\,\mathrm{TLR}(\beta_0^{(k)})\;\leq\; d_Z\,\hat F-\epsilon.
  \end{aligned}
  \end{align}
  By {\cref{eq:schur-marg} applied to the other block of \(\hat\Sigma^{-1}\)}, we have
  \(n(\hat\delta-\delta_n^{(k)})'\hat\Sigma_{\hat\delta}^{-1}(\hat\delta-\delta_n^{(k)})\leq d_Z\,\hat F-\epsilon\). 
  Both \(\gamma_n^{(k)}\) and \(\delta_n^{(k)}\) are thus bounded uniformly in \(k\). At \(\beta_0^{(k)}\), we have the  TLR constraint as
    \begin{align}
     \gamma_n^{(k)}{}'\hat\Sigma_{ZZ}\delta_n^{(k)} \;=\; \beta_0^{(k)}\,\gamma_n^{(k)}{}'\hat\Sigma_{ZZ}\gamma_n^{(k)}.      
    \end{align}
  where the
  left-hand side is bounded uniformly in \(k\).  It follows that
   \(\beta_0^{(k)}\gamma_n^{(k)}{}'\hat\Sigma_{ZZ}\gamma_n^{(k)}=O(1)\). Since,
  by assumption, \(\lvert\beta_0^{(k)}\rvert\to\infty\), we must thus have
  \begin{align}
  \gamma_n^{(k)}{}'\hat\Sigma_{ZZ}\gamma_n^{(k)}\;\conv\;0.
  \end{align}
    Combined with the fact that \(\hat\Sigma_{ZZ}\) is positive definite, this forces \(\gamma_n^{(k)}\to 0\). 
    Substituting into \cref{eq:gamma-bound},  and considering 
  \(k\to\infty\), we have
  \[
  n\,\hat\gamma'\hat\Sigma_{\hat\gamma}^{-1}\hat\gamma \;=\; d_Z\,\hat F\;\leq\; d_Z\,\hat F-\epsilon,
  \]
  This is a contradiction. It follows that \(\liminf_{\lvert\beta_0\rvert\to\infty}d_Z\,\mathrm{TLR}(\beta_0)\geq d_Z\,\hat F\). 
  Combined with
  \cref{eq:tlr-univ-bound}, we  must have that
  \begin{align}\label{eq:tlr-lim-Fhat}
  d_Z\,\mathrm{TLR}(\beta_0)\;\conv\;d_Z\,\hat F
  \end{align}
  as \(\lvert\beta_0\rvert\to\infty\).

  \paragraph*{Critical value limit.}

  By \Cref{prop:bd}, at \(\varrho=\pm 1\), we have \(d_Z\,\mathrm{TLR}(\beta_0)\convd \mathfrak{C}^2_{d_Z}\).
  Hence
  \begin{align}\label{eq:c-at-rho-pm1}
  \lim_{\lvert\varrho\rvert \to 1}c_{1-\alpha_2}(\varrho,\xi)\;=\;\chi^2_{d_Z,1-\alpha_2}
  \end{align}
  for any \(\xi\in\R_+\).
  {The mapping \(\varrho\mapsto c_{1-\alpha_2}(\varrho,\xi)\) is continuous on \([-1,1]\) for any fixed \(\xi\).} Combined with \(\lim_{\beta_0\to \pm \infty}\varrho(\beta_0) = \mp 1\), we obtain for any \(\xi\in \hat \Xi_{1-\alpha_1}\),
  \begin{align}\label{eq:cutoff-limit}
  \lim_{\lvert\beta_0\rvert\to\infty}c_{1-\alpha_2}(\varrho(\beta_0),\xi)=\chi^2_{d_Z,1-\alpha_2}.
  \end{align}
  {The first-stage confidence interval, \(\hat \Xi_{1-\alpha_1}\), is a function of \(\hat S\) only, hence fixed and compact as \(\beta_0\to\pm\infty\) and the convergence in \cref{eq:cutoff-limit} is uniform, hence \(\max_{\xi\in\hat\Xi_{1-\alpha_1}}c_{1-\alpha_2}(\varrho(\beta_0),\xi)\to\chi^2_{d_Z,1-\alpha_2}\).}
  It thus follows that
  \begin{align}
      \lim_{\lvert\beta_0\rvert\to\infty}\1\left[d_Z\mathrm{TLR}(\beta_0) > \max_{\xi \in \hat \Xi_{1-\alpha_1}}c_{1-\alpha_2}(\hat \varrho,\xi)\right] = \1\left[d_Z\hat F > \chi^2_{d_Z,1-\alpha_2}\right]
  \end{align}
  which rejects if and only if \(d_Z\hat F > \chi^2_{d_Z,1-\alpha_2}\).

\end{proof}


\clearpage

\begin{thebibliography}{}

\bibitem[Anderson and Rubin, 1949]{andersonEstimationParametersSingle1949}
Anderson, T.~W. and Rubin, H. (1949).
\newblock Estimation of the parameters of a single equation in a complete system of stochastic equations.
\newblock {\em The Annals of Mathematical Statistics}, 20(1):46--63.

\bibitem[Andrews et~al., 2006]{andrewsOptimalTwoSidedInvariant2006}
Andrews, D. W.~K., Moreira, M.~J., and Stock, J.~H. (2006).
\newblock Optimal two-sided invariant similar tests for instrumental variables regression.
\newblock {\em Econometrica}, 74(3):715--752.

\bibitem[Andrews et~al., 2019]{andrewsWeakInstrumentsInstrumental2019}
Andrews, I., Stock, J.~H., and Sun, L. (2019).
\newblock Weak instruments in instrumental variables regression: Theory and practice.
\newblock {\em Annual Review of Economics}, 11:727--753.

\bibitem[Angrist et~al., 2000]{angristInterpretationInstrumentalVariables2000}
Angrist, J.~D., Graddy, K., and Imbens, G.~W. (2000).
\newblock The interpretation of instrumental variables estimators in simultaneous equations models with an application to the demand for fish.
\newblock {\em The Review of Economic Studies}, 67(3):499--527.

\bibitem[Angrist and Imbens, 1995]{angristTwoStageLeastSquares1995}
Angrist, J.~D. and Imbens, G.~W. (1995).
\newblock Two-stage least squares estimation of average causal effects in models with variable treatment intensity.
\newblock {\em Journal of the American Statistical Association},
  90(430):431--442.

\bibitem[Berger and Boos, 1994]{bergerValuesMaximizedConfidence1994}
Berger, R.~L. and Boos, D.~D. (1994).
\newblock P values maximized over a confidence set for the nuisance parameter.
\newblock {\em Journal of the American Statistical Association},
  89(427):1012--1016.

\bibitem[Bound et~al., 1995]{boundProblemsInstrumentalVariables1995}
Bound, J., Jaeger, D.~A., and Baker, R.~M. (1995).
\newblock Problems with instrumental variables estimation when the correlation between the instruments and the endogenous explanatory variable is weak.
\newblock {\em Journal of the American Statistical Association},
  90(430):443--450.

\bibitem[Chernozhukov et~al.,
  2013]{chernozhukovIntersectionBoundsEstimation2013}
Chernozhukov, V., Lee, S., and Rosen, A.~M. (2013).
\newblock Intersection bounds: Estimation and inference.
\newblock {\em Econometrica}, 81(2):667--737.

\bibitem[Chyn et~al., 2025]{chynExaminerJudgeDesigns2025}
Chyn, E., Frandsen, B., and Leslie, E. (2025).
\newblock Examiner and judge designs in economics: A practitioner's guide.
\newblock {\em Journal of Economic Literature}, 63(2):401--439.

\bibitem[Evdokimov and Koles{\'a}r,
  2018]{evdokimovInferenceInstrumentalVariables2018}
Evdokimov, K.~S. and Koles{\'a}r, M. (2018).
\newblock Inference in instrumental variables analysis with heterogeneous treatment effects.
\newblock Working Paper.

\bibitem[Finlay and Magnusson,
  2009]{finlayImplementingWeakInstrumentRobust2009}
Finlay, K. and Magnusson, L.~M. (2009).
\newblock Implementing weak-instrument robust tests for a general class of instrumental-variables models.
\newblock {\em The Stata Journal}, 9(3):398--421.

\bibitem[Fuller, 1977]{fullerPropertiesModificationLimited1977}
Fuller, W.~A. (1977).
\newblock Some properties of a modification of the limited information estimator.
\newblock {\em Econometrica}, 45(4):939--953.

\bibitem[{Goldsmith-Pinkham} et~al.,
  2025]{goldsmith-pinkhamLeniencyDesignsOperators2025}
{Goldsmith-Pinkham}, P., Hull, P., and Koles{\'a}r, M. (2025).
\newblock Leniency designs: An operator's manual.
\newblock Working Paper arXiv:2511.03572, arXiv.

\bibitem[Hansen, 1982]{hansenLargeSampleProperties1982}
Hansen, L.~P. (1982).
\newblock Large sample properties of generalized method of moments estimators.
\newblock {\em Econometrica}, 50(4):1029--1054.

\bibitem[Hansen, 2005]{hansenTestSuperiorPredictive2005}
Hansen, P.~R. (2005).
\newblock A test for superior predictive ability.
\newblock {\em Journal of Business \& Economic Statistics}, 23(4):365--380.

\bibitem[Heckman et~al., 2006]{heckmanUnderstandingInstrumentalVariables2006}
Heckman, J.~J., Urzua, S., and Vytlacil, E. (2006).
\newblock Understanding instrumental variables in models with essential heterogeneity.
\newblock {\em The Review of Economics and Statistics}, 88(3):389--432.

\bibitem[Imbens and Angrist, 1994]{imbensIdentificationEstimationLocal1994b}
Imbens, G.~W. and Angrist, J.~D. (1994).
\newblock Identification and estimation of local average treatment effects.
\newblock {\em Econometrica}, 62(2):467--475.

\bibitem[Kleibergen, 2002]{kleibergenPivotalStatisticsTesting2002}
Kleibergen, F. (2002).
\newblock Pivotal statistics for testing structural parameters in instrumental variables regression.
\newblock {\em Econometrica}, 70(5):1781--1803.

\bibitem[Koles{\'a}r, 2013]{kolesarEstimationInstrumentalVariables2013}
Koles{\'a}r, M. (2013).
\newblock Estimation in an instrumental variables model with treatment effect heterogeneity.
\newblock Working Paper.

\bibitem[Lee et~al., 2022]{leeValidTRatioInference2022}
Lee, D.~S., McCrary, J., Moreira, M.~J., and Porter, J.~R. (2022).
\newblock Valid t-ratio inference for IV.
\newblock {\em American Economic Review}, 112(10):3260--3290.

\bibitem[Lee et~al., 2023]{leeWhatWhenYou2023}
Lee, D.~S., McCrary, J., Moreira, M.~J., Porter, J.~R., and Yap, L. (2023).
\newblock What to do when you can't use '1.96' confidence intervals for IV.
\newblock NBER Working Paper 31893.

\bibitem[Loh, 1985]{lohNewMethodTesting1985}
Loh, W.-Y. (1985).
\newblock A new method for testing separate families of hypotheses.
\newblock {\em Journal of the American Statistical Association},
  80(390):362--368.

\bibitem[McCloskey,
  2017]{mccloskeyBonferronibasedSizecorrectionNonstandard2017}
McCloskey, A. (2017).
\newblock Bonferroni-based size-correction for nonstandard testing problems.
\newblock {\em Journal of Econometrics}, 200(1):17--35.

\bibitem[Mogstad and Torgovitsky,
  2024]{mogstadInstrumentalVariablesUnobserved2024}
Mogstad, M. and Torgovitsky, A. (2024).
\newblock Instrumental variables with unobserved heterogeneity in treatment effects.
\newblock NBER Working Paper 32927.

\bibitem[Mogstad et~al., 2021]{mogstadCausalInterpretationTwoStage2021}
Mogstad, M., Torgovitsky, A., and Walters, C.~R. (2021).
\newblock The causal interpretation of two-stage least squares with multiple instrumental variables.
\newblock {\em American Economic Review}, 111(11):3663--3698.

\bibitem[Moreira, 2003]{moreiraConditionalLikelihoodRatio2003}
Moreira, M.~J. (2003).
\newblock A conditional likelihood ratio test for structural models.
\newblock {\em Econometrica}, 71(4):1027--1048.

\bibitem[Moreira, 2009]{moreiraTestsCorrectSize2009}
Moreira, M.~J. (2009).
\newblock Tests with correct size when instruments can be arbitrarily weak.
\newblock {\em Journal of Econometrics}, 152(2):131--140.

\bibitem[Nelson and Startz, 1990a]{nelsonDistributionInstrumentalVariables1990}
Nelson, C.~R. and Startz, R. (1990a).
\newblock The distribution of the instrumental variables estimator and its t-ratio when the instrument is a poor one.
\newblock {\em The Journal of Business}, 63(1):S125--S140.

\bibitem[Nelson and Startz, 1990b]{nelsonFurtherResultsExact1990}
Nelson, C.~R. and Startz, R. (1990b).
\newblock Some further results on the exact small sample properties of the instrumental variable estimator.
\newblock {\em Econometrica}, 58(4):967--976.

\bibitem[Romano et~al., 2014]{romanoPracticalTwoStepMethod2014}
Romano, J.~P., Shaikh, A.~M., and Wolf, M. (2014).
\newblock A practical two-step method for testing moment inequalities.
\newblock {\em Econometrica}, 82(5):1979--2002.

\bibitem[Sargan, 1958]{sarganEstimationEconomicRelationships1958}
Sargan, J.~D. (1958).
\newblock The estimation of economic relationships using instrumental variables.
\newblock {\em Econometrica}, 26(3):393--415.

\bibitem[Serfling, 1980]{ApproximationTheoremsMathematical1980}
Serfling, R.~J. (1980).
\newblock {\em Approximation theorems of mathematical statistics}.
\newblock Wiley, New York, 1st edition.

\bibitem[Sigstad, 2026]{sigstadMonotonicityJudgesEvidence2026}
Sigstad, H. (2026).
\newblock Monotonicity among judges: Evidence from judicial panels and consequences for judge IV designs.
\newblock {\em American Economic Review}, 116(1):189--208.

\bibitem[Silvapulle, 1996]{silvapulleTestPresenceNuisance1996}
Silvapulle, M.~J. (1996).
\newblock A test in the presence of nuisance parameters.
\newblock {\em Journal of the American Statistical Association},
  91(436):1690--1693.

\bibitem[Staiger and Stock, 1997]{staigerInstrumentalVariablesRegression1997}
Staiger, D. and Stock, J.~H. (1997).
\newblock Instrumental variables regression with weak instruments.
\newblock {\em Econometrica}, 65(3):557--586.

\bibitem[Stern and Wolkowicz, 1995]{sternIndefiniteTrustRegion1995}
Stern, R.~J. and Wolkowicz, H. (1995).
\newblock Indefinite trust region subproblems and nonsymmetric eigenvalue perturbations.
\newblock {\em SIAM Journal on Optimization}, 5(2):286--313.

\bibitem[{Van de Sijpe} and Windmeijer,
  2023]{vandesijpePowerConditionalLikelihood2023}
{Van de Sijpe}, N. and Windmeijer, F. (2023).
\newblock On the power of the conditional likelihood ratio and related tests for weak-instrument robust inference.
\newblock {\em Journal of Econometrics}, 235(1):82--104.

\bibitem[Vytlacil, 2002]{vytlacilIndependenceMonotonicityLatent2002a}
Vytlacil, E. (2002).
\newblock Independence, monotonicity, and latent index models: An equivalence result.
\newblock {\em Econometrica}, 70(1):331--341.

\bibitem[Yap, 2025]{yapInferenceManyWeak2025}
Yap, L. (2025).
\newblock Inference with many weak instruments and heterogeneity.
\newblock Working Paper.

\end{thebibliography}

\clearpage

\begin{thebibliography}{}

\bibitem[Johnson et~al., 1995]{johnsonContinuousUnivariateDistributions1995}
Johnson, N.~L., Kotz, S., and Balakrishnan, N. (1995).
\newblock {\em Continuous univariate distributions, Volume 2}.
\newblock Wiley, New York, 2nd edition.

\bibitem[Lehmann and Romano, 2005]{lehmannTestingStatisticalHypotheses2005}
Lehmann, E.~L. and Romano, J.~P. (2005).
\newblock {\em Testing statistical hypotheses}.
\newblock Springer Texts in Statistics. Springer, New York, NY, 3rd edition.

\bibitem[Stern and Wolkowicz, 1995]{sternIndefiniteTrustRegion1995}
Stern, R.~J. and Wolkowicz, H. (1995).
\newblock Indefinite trust region subproblems and nonsymmetric eigenvalue perturbations.
\newblock {\em SIAM Journal on Optimization}, 5(2):286--313.

\end{thebibliography}
\end{document}